\documentclass[prd,aps,floats,floatfix,eqsecnum,nofootinbib]{revtex4}
\usepackage{amsmath,amssymb,verbatim,epsfig,graphicx,rotating}
\usepackage{psfrag}
\newcommand{\be}{\begin{equation}}
\newcommand{\ee}{\end{equation}}
\newcommand{\bea}{\begin{eqnarray}}
\newcommand{\eea}{\end{eqnarray}}
\newcommand{\vk}{\vec{k}}
\newcommand{\vp}{\vec{p}}
\newcommand{\td}{ \delta}

\begin{document}
\title{Clustering properties of a sterile neutrino dark matter candidate.}
\author{\bf D. Boyanovsky }
\email{boyan@pitt.edu}
\affiliation{ Department of Physics and Astronomy,
University of Pittsburgh, Pittsburgh, Pennsylvania 15260,
USA.}

\date{\today}

\begin{abstract}
The clustering properties of sterile neutrinos are studied within a
simple extension of the minimal standard model, where these
neutrinos are produced via the decay of a gauge singlet scalar. The
distribution function after decoupling is
strongly out of equilibrium and features an enhancement at small
comoving momentum $\propto 1/\sqrt{p}$. Dark matter abundance and
phase space density constraints from
dwarf spheroidal galaxies constrain the mass in the $\mathrm{keV}$
range consistent with a Yukawa coupling to a gauge singlet with mass
and vacuum expectation value in the range $\sim 100\,\mathrm{GeV}$
and a decoupling temperature of this order. The dark matter
transfer function and power spectrum are obtained from the solution of
the non-relativistic Boltzmann-Vlasov equation in the matter
dominated era. The small momentum enhancement of the non-equilibrium
distribution function leads to long range memory of gravitational
clustering and a \emph{substantial enhancement of the power spectrum
at small scales as compared to a thermal relic  or sterile neutrino
produced via non-resonant mixing with active neutrinos}. The scale
of suppression of the power spectrum for a sterile neutrino with
$m\sim \mathrm{keV}$ produced by scalar decay that decouples at
$\sim 100 \mathrm{GeV}$ is $\lambda \sim 488 \,\mathrm{kpc}$. At large scales   $T(k)\sim 1-C \,  k^2/k^2_{fs}(t_{eq})  +\cdots$ with $C \sim \mathrm{O}(1)$. At small scales $65\,\mathrm{kpc} \lesssim \lambda \lesssim 500\,\mathrm{kpc}$  corrections to the fluid description and memory of gravitational
  clustering become important, and we find $T(k) \simeq 1.902\,e^{-k/k_{fs}(t_{eq})}$, where
$k_{fs}(t_{eq}) \sim 0.013/\mathrm{kpc}$ is the free streaming wavevector at matter-radiation equality. The enhancement of power at small scales may provide a possible relief to the tension between the
constraints from X-ray and Lyman-$\alpha$ forest data.

\end{abstract}

\maketitle

\section{Introduction}

In the \emph{concordance} $\Lambda\mathrm{CDM}$ standard cosmological
model  dark matter (DM) is composed of primordial particles which are
cold and collisionless\cite{primack}.  In this  cold dark matter (CDM) scenario
  structure formation proceeds in a hierarchical ``bottom up''
approach: small scales become non-linear and collapse first and
their merger and accretion leads to structure on larger scales. CDM particles
feature negligible small velocity dispersion  leading to a power spectrum that favors small
scales. In this hierarchical scenario, dense clumps that survive the
merger process form satellite galaxies.

Large scale simulations seemingly yield  an overprediction of
satellite galaxies\cite{moore2} by almost an order of magnitude
larger than
 the number of satellites that have been observed in Milky-Way
sized galaxies\cite{kauff,moore,moore2,klyp,will}. Simulations
within the $\Lambda$CDM paradigm also yield a density profile in
virialized (DM) halos that
increases monotonically towards the
center\cite{frenk,dubi,moore2,bullock,cusps} and features a cusp, such as the
Navarro-Frenk-White (NFW) profile\cite{frenk} or more general
  central density profiles $\rho(r) \sim r^{-\beta}$ with
$1\leq \beta \lesssim 1.5$\cite{moore,frenk,cusps}. These density
profiles accurately describe clusters of galaxies but there is an
accumulating body of observational
evidence\cite{dalcanton1,van,swat,gilmore,salucci,battaglia,cen,wojtak}
that seem to indicate that the     central regions of (DM)-dominated
dwarf spheroidal satellite (dSphs) galaxies
 feature smooth cores instead of cusps as predicted by (CDM). More
 recently\cite{cen} a ``galaxy size'' problem has been reported,
 where large scale simulations at $z=3$ yield galaxies that are too
 small, this problem has been argued to be related to that of the
 missing dwarf galaxies.

Warm dark matter (WDM) particles were
invoked\cite{mooreWDM,turokWDM,avila} as possible solutions to these
discrepancies both in the over abundance of satellite galaxies and
as a mechanism to smooth out  the cusped   density profiles
predicted by (CDM) simulations into the  cored profiles that fit the
observations in   (dShps). (WDM) particles feature a range of
velocity dispersion in between the (CDM) and hot dark matter (HDM)
leading to free streaming scales that smooth  out small scale
features and could be consistent with core radii of the (dSphs). If
the free streaming scale of these particles is  smaller than the
scale of galaxy clusters, their large scale structure properties are
indistinguishable from (CDM) but may affect the  \emph{small} scale
power spectrum\cite{bond} so as to provide an explanation of the
smoother inner profiles of (dSphs), fewer satellites and the size of
galaxies at $z=3$\cite{cen}.

Sterile neutrinos with masses $\sim \mathrm{keV}$ may be suitable
(WDM) candidates\cite{dw,este,shapo,kusenko,kusepetra,petra}. The
main property that is relevant for structure formation  of any dark
matter candidate is its distribution function after
decoupling\cite{hogan,coldmatter}, which depends on the production
mechanism and the (quantum) kinetics of its evolution from
production to decoupling. There is a variety of mechanisms of
sterile neutrino production\cite{dw,colombi,este,shapo,kuse2}, and
mixing between sterile and active neutrinos can be one of
them\cite{dw,colombi,este}. There is considerable tension between
the X-ray\cite{xray} and Lyman-$\alpha$ forest\cite{lyman,lyman2}
data if sterile neutrinos are produced via the Dodelson-Widrow
(DW)\cite{dw} non-resonant mixing  mechanism, leading to the
suggestion\cite{palazzo} that these cannot be the dominant (DM)
component. Constraints from the Lyman-$\alpha$ forest spectra are
particularly important because of its sensitivity to the suppression
of the power spectrum by free-streaming in the linear
regime\cite{lyman,lyman2}. The most recent constraints from the
Lyman-$\alpha$ forest\cite{lyman2} improve upon previous ones, but
rely on the Dodelson-Widrow\cite{dw} model for the distribution
function of sterile neutrinos, leaving open the possibility of
evading these tight constraints with non-equilibrium distribution
functions from other production mechanisms.

The gravitational clustering properties of collisionless (DM) in the
linear regime are described by the   power spectrum of gravitational
perturbations. Free streaming of collisionless (DM)  leads to a
suppression of the transfer function on length scales smaller than
the free streaming scale via Landau
damping\cite{bond,bondszalay,kt}. This scale is   determined by the
decoupling temperature, the particle's mass and the distribution
function at decoupling\cite{freestream}.

In this article we study the gravitational clustering properties of
sterile neutrinos as (WDM) candidates within a model in which these
are produced via the decay of a gauge singlet scalar. Such model has
been advocated recently in ref.\cite{kusepetra,petra,kuse2}, and is
based on a   phenomenologically appealing extension of the minimal
standard model\cite{shapo} consistent with the observed neutrino
masses and mixing. This model also shares many features in common
with  models of gravitino production\cite{rubakov}, hence it
provides a   viable extension of the standard model to study in
detail the production
 and clustering  properties of potential (WDM) candidates. In this model sterile neutrinos   decouple
a temperatures much larger ($\sim 100 \,\mathrm{GeV}$ )  than in the
Dodelson-Widrow scenario ($\sim 150 \,\mathrm{MeV}$)\cite{dw},
therefore they are   \emph{colder} at matter-radiation equality (and
today), being dubbed, for this reason,  ``chilled'' neutrinos in
refs.\cite{kusepetra,petra}.  Clustering properties of active neutrinos in non-standard
cosmology, for example quintessence have been reported in ref.\cite{pettorino}.

A program that yields a quantitative assessment of a particle
physics candidate for (DM) in the linear regime implements the
following steps:

\begin{itemize}
\item{Establish the quantum kinetic equations that describe the production of these particles and
follows the evolution of their distribution function through their decoupling from the cosmological
plasma.}

\item{The distribution function after decoupling becomes the \emph{unperturbed} distribution, which
determines the abundance,   the primordial phase space
densities\cite{hogan,coldmatter} and free streaming
lengths\cite{freestream}. The generalized  Tremaine-Gunn\cite{TG}
constraints  obtained in\cite{coldmatter} in combination with the
recent photometric observations of the phase space densities of
(dSphs) combined with the  DM  abundance lead to
  bounds on the mass, couplings and decoupling
temperature\cite{coldmatter}.   }

\item{The unperturbed distribution function is input in the Boltzmann-Vlasov equation for
density and gravitational perturbations\cite{ma}.  The solution of
which yields the transfer function, and the power spectrum. }

\end{itemize}

We follow this program in the   model of sterile neutrino production
  proposed in
refs.\cite{shapo,kusepetra,petra,kuse2}. In principle, in order to
obtain the transfer function and the power spectrum of density
perturbations the coupled set of Boltzmann equations for baryons,
photons, dark matter and gravitational perturbations must be
solved\cite{ma,dodelson}.
 Photons and baryons are coupled by Thompson scattering and dark matter only couples to the gravitational
  perturbations that are sourced by all the components. In practice this is a computationally daunting task because
popular codes\cite{lewis} based on the set of coupled Boltzmann equations for photons, baryons and dark matter\cite{ma}
 need to be modified to input arbitrary non-equilibrium distribution functions, masses and
couplings.

Recently a simple analytic framework to obtain the dark matter transfer
function, and consequently the power spectrum during matter domination
has been presented\cite{gildan}. The main premise of this
formulation is that the contribution from baryons and photons
modifies the  DM  transfer function at most by a few percent\cite{EH,HS}
  during matter domination and that a preliminary robust assessment of the clustering properties
of a DM candidate can be systematically established by neglecting in
first approximation the contribution from baryons and photons. The influence of baryons
on the DM power spectrum is more prominent on the scale of baryon acoustic oscillations (BAO), corresponding
to the scale of the sound horizon at recombination, or   $\sim 150\,\mathrm{Mpc}$ today\cite{eisenstein}. On this
scale the DM power spectrum does not distinguish between (CDM) or (WDM), and at smaller scales,
of interest for the satellite and cusp problems, the (BAO) features are not prominent and can be
safely neglected.

 The
main ingredient to study the (DM) transfer function in absence of baryons is the non-relativistic Boltzmann-Vlasov equation
for  DM  density and gravitational perturbations. The
non-relativistic limit is warranted for particles that decoupled
early and became non-relativistic prior to matter-radiation equality
and for perturbations
 that entered the horizon prior to matter-radiation equality, these describe all the relevant scales for
  structure formation.

 The method developed in ref.\cite{gildan} yields a simple analytic approximation to the transfer
 function that is remarkably accurate in a wide range of scales relevant to structure formation. An
 important ingredient is a non-local kernel that depends on the unperturbed distribution function
 of the decoupled particles. This kernel describes memory of gravitational clustering and is a
 \emph{correction to the fluid description} which become important at small scales. Distribution functions that feature larger support at small momentum
 yield longer range memory kernels thereby enhancing the transfer function at small scales\cite{gildan}placing  greater importance on non-equilibrium aspects
  of the distribution function.

    In this article we study the clustering
    properties of sterile neutrinos in the model proposed in
    references\cite{kusepetra,petra,shapo,kuse2} by implementing all
    the steps described above, from obtaining and solving the
    quantum kinetic equation for production that establishes the
    distribution function at decoupling, narrowing the range of
    parameters, masses and couplings with the observational constraints from (DM) abundance
    and coarse grained phase space densities of (DM) dominated (dSphs)\cite{gilmore,coldmatter}, and   solving
    the Boltzmann-Vlasov equation obtaining the transfer function
    and power spectrum which we compare to the case of thermal
    relics or Dodelson-Widrow-type\cite{dw} distribution functions.

\vspace{2mm}

{\bf Summary of results:}

The production of  sterile neutrinos of $m\sim \mathrm{keV}$ via the decay of scalar gauge singlet with $M\sim 100 \,\mathrm{GeV}$    leads to decoupling at a temperature $\sim 100 \,\mathrm{GeV}$ and a distribution function that is strongly out of equilibrium and  behaves as $1/\sqrt{p}$ for small comoving momenta $p$.

The constraints from (DM) abundance and coarse-grained phase space density from the latest compilation
of photometric data from (dSphs) lead to a narrow window in the $\mathrm{keV}$ range for the value
of the mass of the sterile neutrino, consistently with the phenomenologically motivated extension beyond the standard model studied.

The (DM) transfer function and power spectrum are obtained from the solution of the non-relativistic
Boltzmann-Vlasov equation for (DM) density and gravitational perturbations during matter domination.
We implement a simple analytic approximate method\cite{gildan} to obtain the density and gravitational perturbations and the transfer function that is remarkably accurate
in a wide range of cosmologically relevant scales, as confirmed by the exact solution. This approach yields a wealth of information    that relates the small scale
behavior of the transfer function to the range of memory of gravitational clustering, which is determined by the small (comoving) momentum region of the distribution function.

The enhancement of the non-equilibrium distribution at small
momentum leads to a long range memory of gravitational clustering
and slower fall off of the free-streaming solution. Both features lead to
an enhancement of the transfer function and power spectrum at small
scales.

We compare the transfer function and power spectrum from sterile neutrinos produced via gauge
singlet decay to that of relativistic fermions decoupled in local thermodynamic equilibrium (LTE) (thermal relic) and sterile neutrinos produced by non-resonant mixing with active neutrinos \emph{a la} Dodelson-Widrow\cite{dw}.
 Thermal relics and (DW)-produced sterile neutrinos feature the \emph{same} transfer function for similar ratios  of the decoupling temperature to mass.

The transfer function and power spectrum for sterile neutrinos produced by scalar decay is \emph{substantially enhanced} with respect to that of (DW)-sterile neutrinos (and thermal relics) at small scales $ \lambda \lesssim 500 \,\mathrm{kpc}$. Whereas for (DW)-sterile neutrinos with  $m\sim \mathrm{keV}$ the transfer function is suppressed on scales $\lambda \lesssim 900 \,\mathrm{kpc}$, the
scale of suppression for  $m\sim \mathrm{keV}$ sterile neutrinos produced by scalar decay at a
scale $\sim 100\,\mathrm{GeV}$ is $\lambda \lesssim 488 \,\mathrm{kpc}$ with a large enhancement of power at smaller scales.

For sterile neutrinos produced by scalar decay we find the following   behavior for the transfer function: at long wavelengths,  \be T(k) \simeq 1- C \bigg(\frac{k}{k_{fs}(t_{eq})}\bigg)^2 + \cdots  ~~;~~ k\ll k_{fs}(t_{eq}) \ee with $C \sim \mathrm{O}(1)$,   and at small scales  where the corrections to the fluid description and the memory of gravitational clustering becomes important   \be T(k) \simeq 1.902\,e^{-k/k_{fs}(t_{eq})}~~;~~k \geq k_{fs}(t_{eq})\,   \ee valid for scales $65\,\mathrm{kpc} \lesssim \lambda \lesssim 500\,\mathrm{kpc}$ where $k_{fs}(t_{eq}) $ is the free streaming wavevector at matter radiation
 equality.  For $m\sim \mathrm{keV}$ and decoupling temperature $\sim 100\,\mathrm{GeV}$ we obtain
$k_{fs}(t_{eq}) \sim 0.013/\mathrm{kpc}$.

The smaller suppression scale may relieve the
tension between the X-ray\cite{xray} and Lyman-$\alpha$ forest\cite{lyman,lyman2} data and may provide the
necessary enhancement of power at small scales to smooth out the inner profile of  (dSphs).

\section{The model}\label{sec:model}

We study the model presented in
references\cite{kusepetra,kuse2,shapo,petra} as an extension of the minimal
standard model with only one sterile neutrino, however,  including more species is straightforward. The Lagrangian density is given by

\be \mathcal{L} = \mathcal{L}_{SM}+ \frac{1}{2}\partial_\mu \chi \partial^\mu\chi -\frac{M^2}{2} \chi^2+ i
\overline{\nu}
{\not\!{\partial}} \nu - \frac{Y }{2} \chi \,\overline{\nu}^c \nu -\frac{m }{2}\, \overline{\nu}^c \nu
 -y_{\alpha }H^\dagger \overline{L}_\alpha \,\nu   - V(H^\dagger H;\chi)+\mathrm{h.c.} \label{lagra}\ee
 where $\mathcal{L}_{SM}$ is the standard model Lagrangian, $L_{\alpha};\alpha=1,2,3$ are the standard
 model $SU(2)$ lepton doublets, $\nu $ is a singlet sterile neutrino with a (Majorana) mass $m $, a
 real scalar $\chi$ with Yukawa coupling $Y $ to the sterile
neutrino which in turn is Yukawa coupled to the active neutrinos via
the Higgs doublet $H$, thereby building a see-saw mass matrix in
terms of the vacuum expectation of this
doublet\cite{kusepetra,kuse2,shapo}.

As discussed in detail in ref.\cite{coldmatter} abundance and phase
space density constraints from (dSphs) indicate that the mass of
suitable (WDM) candidates must be in the $\mathrm{keV}$ range, which
leads to considering
  the vacuum expectation value and mass $M$ of the singlet scalar $\chi$
   in the range $\sim 100\,\mathrm{GeV}$ as discussed in references\cite{kusepetra,petra,kuse2}.
    If the sterile neutrino mass $m \sim \mathrm{keV}$ results from
  the vacuum expectation value $\langle \chi \rangle \sim 100\,\mathrm{GeV}$ then the Yukawa coupling
  $Y \sim 10^{-8}$.

  The results of the study here demonstrate that  this range of parameters yields a consistent description
  of sterile neutrinos as a suitable (DM) candidate in this model.

\subsection{Decoupling out of equilibrium}

Non-resonant active-sterile mixing leads to sterile neutrino
production via the Dodelson-Widrow mechanism\cite{dw} with a
decoupling temperature near the QCD scale\cite{dw,este} $\sim 150
\,\mathrm{MeV}$. In this scenario the distribution function at
decoupling is of the form\cite{dw} \be f_{dw}(P_f;T) =
\frac{\beta}{e^y+1} ~~;~~y=\frac{P_f}{T} ~~;~~ 0 < \beta \leq 1
\label{fdw} \ee where $P_f$ is the physical momentum.  For sterile
neutrinos produced by non-resonant mixing with active
neutrinos\cite{dw} $\beta \propto \theta^2_m$ where $\theta_m
\lesssim 10^{-2}$\cite{xray,este} is the mixing angle. A fermionic
relic decoupled in local thermodynamic equilibrium (LTE) while
relativistic corresponds to $\beta =1$. This general  type of
distribution function with a suppression factor $\beta$   has been
used in the Lyman-$\alpha$ forest analysis\cite{lyman2}.

In this article we will neglect this production mechanism and focus
on the production via the decay of the gauge singlet scalar field
$\chi$ which, as discussed in references\cite{kusepetra,kuse2,petra}
lead to ``colder'' relics. However, we will compare the clustering
properties of the distribution obtained via this mechanism and that
of   relics that decoupled with the generalized distributions
(\ref{fdw}), postponing to another study the complete kinetic
description that accounts for both processes. Since these production
mechanisms are effective at widely different scales ($\sim
100\,\mathrm{GeV}$ for $\chi \rightarrow
\overline{\nu}\nu$-decay\cite{kusepetra,kuse2,petra}, vs. $\sim
150\,\mathrm{MeV}$ for Dodelson-Widrow\cite{dw,este}) we expect that
possible corrections from mixing will be subleading and certainly so
for the small $y$ region of interest.  A more complete study is
forthcoming.

We  consider the case in which the Yukawa coupling $Y \ll 1$ and
$M\gg m$ and assume that the scalar field $\chi$ is strongly coupled
to the plasma and is in (LTE) with a Bose-Einstein distribution
function\footnote{The case where the scalar is out of equilibrium has been considered
 in\cite{kusepetra,petra}.} \be N_k = \frac{1}{e^{\Omega_k(t)/T(t)}-1} ~~;~~
\Omega_k(t) = \sqrt{ \frac{k^2}{a^2(t)}+M^2} \label{Nofk}\ee where
$k$ is a comoving wavevector and  \be T(t) = \frac{T_{0}}{a(t)}
\,\label{Toft}\ee where $T_0$ would be the temperature of the plasma
\emph{today}. During the radiation dominated era the Hubble expansion rate is given by\cite{kt} \be H(t)
\simeq 1.66 g^\frac{1}{2}(t) \frac{T^2(t)}{M_{pl}} \label{Hoft}\ee
where $g(t)$ is the effective number of relativistic degrees of
freedom.

The details leading to the quantum kinetic equation for the
production of sterile neutrinos via the decay of the scalar field
$\chi$ are given in the appendix and the final result   is given by
eqn. (\ref{rateeqn}).

For $Y^2\ll1$ we expect that neutrinos will decouple early and their
distribution function will freeze-out with $n_p, \overline{n}_p \ll
1$. This expectation will be confirmed below self-consistently from
the solution of the kinetic equation. Neglecting the neutrino
population buildup in the kinetic equation (\ref{rateeqn}), namely
setting $n_p = \overline{n}_q =0$,   neglecting   terms of order
$m^2/M^2 \ll 1$, taking the scalar field to be in LTE and replacing
the momenta in (\ref{rateeqn}) by their physical values, we find
from (\ref{rateeqn}) \be \frac{d\,n(p;\,t)}{dt} = \frac{Y^2
M^2\,T(t)}{8\pi\,P_f(t)\, \omega_{p}(t)}
\ln\Bigg[\frac{1-e^{-(\omega_+(t)+\omega_p(t))/T(t)}}{1-e^{-(\omega_-(t)+\omega_p(t))/T(t)}}
\Bigg] \label{QKE}\ee where $P_f(t)=p/a(t)$ is the physical
momentum, $p$ is the comoving momentum and \bea  \omega_p(t) &  =
 & \sqrt{\frac{p^2}{a^2(t)}+m^2} \label{omepoft}\\ \omega_{\pm}(t) &
 = & \sqrt{  q^2_{\pm}(t)  +m^2} \label{omepm}\eea where
 $q_{\pm}(t)$  are given by eqn. (\ref{roots}) in the appendix in
 terms of the corresponding physical momenta. These values are determined by the kinematic
 thresholds for scalar decay.

 We anticipate self-consistently,  that for $Y\ll1$ neutrinos   decouple at
 temperatures $T_d \gg m$, namely when they are still relativistic, therefore
 we can safely neglect terms of order $m^2/T^2(t) < m^2/T^2_d \ll
 1$. Under this assumption (to be confirmed self-consistently
 below),  using $P_f(t)/T(t) = p/T_0$ and neglecting terms of
 order $m^2/M^2 \ll1$,  the kinetic equation above simplifies
 considerably. It proves convenient to use the dimensionless variables  \be \tau =
 \frac{M}{T(t)} \label{tau} \ee with \be \frac{d\tau}{dt} = \tau H(t) \label{dtau}\ee and \be y = \frac{p}{T_{0}} \label{y}
 \ee   leading to the following
 form of the quantum kinetic equation, \be
 \frac{dn(y\,;\tau)}{d\tau} = \Lambda(\tau) \Big(\frac{\tau}{y}\Big)^2
 \ln\Bigg[\frac{1-e^{-M^2y/m^2}}{1-e^{-y-\tau^2/4y}} \Bigg] \label{dndtau}\ee
 where we have introduced\be \Lambda(\tau) = \frac{Y^2}{8\pi(1.66 \, g^\frac{1}{2}(t))} \Big(\frac{M_{Pl}}{M}\Big) \label{Lambdaoft}\ee
 and the  effective number of relativistic degrees of freedom depends on time through the temperature.

 Under the assumption $M\gg m$, for example taking  $M \sim 100 \,\mathrm{GeV},m \sim
 \mathrm{keV}$,
  the exponential in the numerator inside the logarithm can be neglected for all $y \gg m^2/M^2 \sim 10^{-16}$. Although we
 are interested in the small momentum region of the distribution function (small $y$), the phase space
 suppression for small momentum in the integrals of the distribution function entails that we can safely neglect the contributions of such small values
 of $y$. This argument will be confirmed explicitly  below. Therefore we
 safely neglect the numerator inside the logarithm in (\ref{dndtau}).
 In order integrate the rate equation (\ref{dndtau}) we must furnish $g(t)$. In the Standard Model this function is approximately constant in large intervals and
   features sharp variations in the regions of temperature when relativistic degrees of freedom either
   decay, annihilate or become non-relativistic\cite{kt}.

    We will \emph{assume} that $g(t)$ is approximately constant in the region of
(large) temperature  before and during decoupling, replacing the
value of $g$ by its average $\overline{g}$ over the range in which
the rate (\ref{dndtau}) is appreciable. The numerical analysis
presented below justifies this assumption for a wide range of $y$.
Therefore we replace \be \Lambda(\tau) \simeq \Lambda =
\frac{Y^2}{8\pi(1.66 \, \overline{g}^\frac{1}{2})}
\Big(\frac{M_{Pl}}{M}\Big) \,. \label{Lambdave}\ee With these
approximations and assumptions, eqn. (\ref{dndtau}) is equivalent to
the quantum kinetic equation obtained in
references\cite{kusepetra,shapo,rubakov} and can be integrated
exactly. We obtain, \be n(y;\tau) = \Lambda \Bigg\{2\sqrt{\pi}~
\frac{g_{\frac{5}{2}}( y)}{y^\frac{1}{2}}+ \frac{\tau^3}{3y^2}
\ln\Big[\frac{1}{1-e^{-y-\tau^2/4y}}\Big] - \frac{8}{3y^2}
\sum_{n=1}^\infty \frac{e^{-ny}}{n^\frac{5}{2}}
~\Gamma\Big[\frac{5}{2},\frac{n\tau^2}{4y}\Big]\Bigg\}\label{nofytau}
\ee where\footnote{Surprisingly the function $g_{\frac{5}{2}}(z)$
also
 determines the equation of state of the ideal non-relativistic
Bose gas.} \be g_{\frac{5}{2}}( y) = \sum_{n=1}^\infty
\frac{e^{-ny}}{n^\frac{5}{2}}~~;~~g_{\frac{5}{2}}(0)=
\zeta\Big(\frac{5}{2}\Big)=1.342\cdots \ee and $\Gamma[a,b]$ is the
incomplete Gamma function.

The rate (\ref{dndtau}) vanishes as $\tau \rightarrow 0$, reaches a
maximum and falls-off exponentially as $\tau/4y \rightarrow \infty$.
The maximum rate is larger for \emph{smaller} values of $y$,  this feature translates into an
\emph{enhancement} of the distribution function at small momenta,
which will be at the heart of the important aspects of clustering
studied   in section (\ref{sec:TF}).

The asymptotic behavior of    the distribution function (\ref{nofytau}) for $\tau/4y \gg 1$ is given by
\be n(y;\tau)   \stackrel{\tau^2/4y \gg 1}{\simeq} \Lambda \Bigg[ 2 \sqrt{\pi}~
\frac{g_{\frac{5}{2}}(y)}{y^\frac{1}{2}}+   \frac{\tau^3}{3y^2}e^{-y} \,
 e^{-\tau^2/4y}-\frac{ \tau^3}{3y^\frac{7}{2}}\sum_{n=1}^\infty \frac{e^{-ny}\,e^{-n\tau^2/4y}}{n} \Bigg]
 \label{asyn}\ee

\begin{figure}[ht!]
\begin{center}
\includegraphics[height=2in,width=3in,keepaspectratio=true]{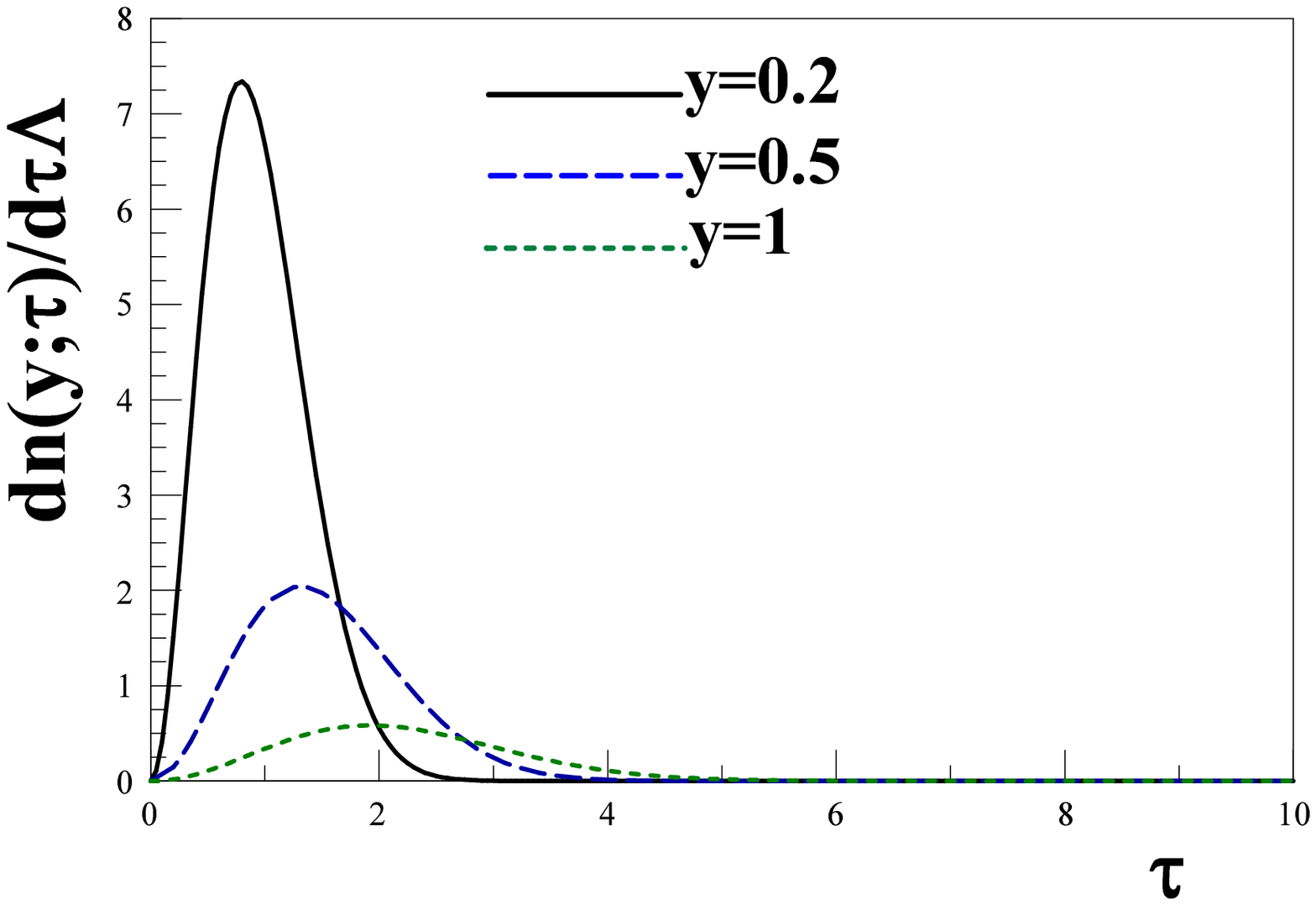}
\includegraphics[height=2in,width=3in,keepaspectratio=true]{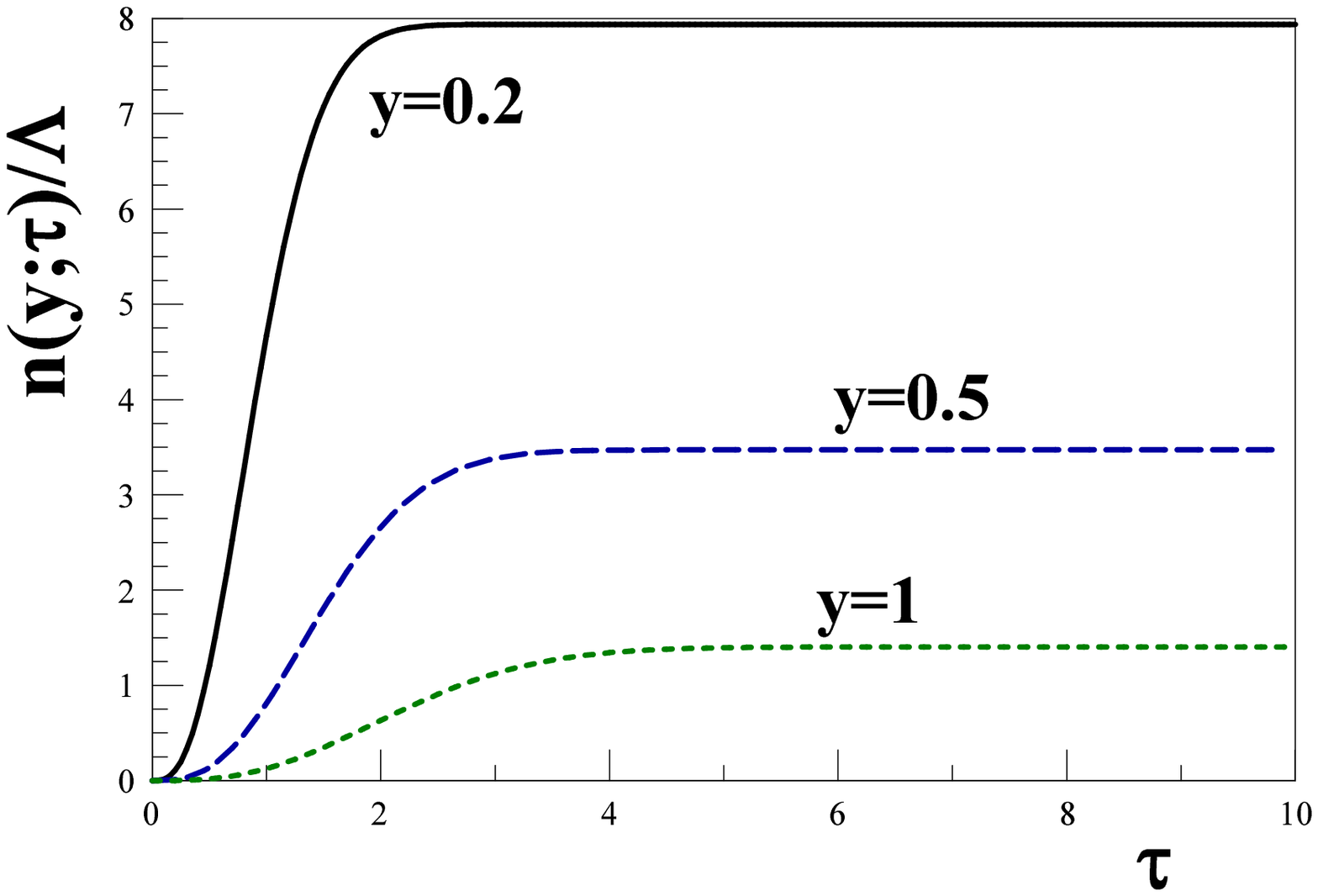}
\caption{Left panel: $dn(y;\tau)/d\tau \Lambda$ for
$y=0.2,0.5,1$. Right panel $n(y;\tau)/\Lambda$ for the
same values of $y$ .} \label{fig:population}
\end{center}
\end{figure}

Figure (\ref{fig:population}) displays the rate (\ref{dndtau}) (left
panel) and the distribution function (\ref{nofytau}) (right panel)
as a function of $\tau$ for several values of $y$. For
  $y \ll 1 $ the distribution function is largest,  reaching its
asymptotic value for $\tau \lesssim  1$, whereas for large values $y \gg
1$ the distribution function is strongly suppressed and reaches its
asymptotic form much later (see fig.\ref{fig:population}).

Hence, for small $y$, the region of distribution function most
relevant for small scale structure formation\cite{gildan},
decoupling occurs fairly fast, at a decoupling temperature $\sim M
\gg m$ thus justifying the assumption of  a constant $g(t)$ and $m/T_d \ll 1$ with $T_d=T_0$.

As discussed above and in ref.\cite{gildan}, the dark matter
transfer function depends more sensitively on the \emph{small}
momentum (small $y$) region of the distribution function, the above
analysis shows that for $y<1$ the distribution function freezes out
on a ``time'' scale $\tau \simeq 1$, namely a temperature scale
$T(t_f) \sim M$, where $t_f$ is the ``freeze-out'' time. Hence if
the effective number of relativistic degrees of freedom $g(t)$
varies smoothly within the temperature range in which the population
freezes-out $\sim M$ the approximation (\ref{Lambdave}) is justified
and reliable. Therefore we take the value $\overline{g}$ as the
effective number of relativistic degrees of freedom at decoupling,
for example for a freeze-out temperature $T(t_f) \sim M \sim
100~\mathrm{GeV}$ in the standard model  $\overline{g} \sim
100$\cite{kt}.

\begin{figure}[ht!]
\begin{center}
\includegraphics[height=2in,width=3in,keepaspectratio=true]{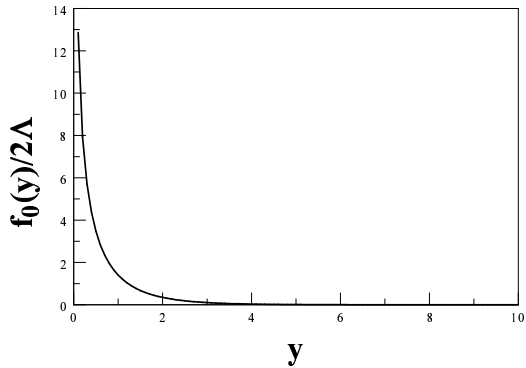}
\caption{ The distribution function $f_0(y)=n(y;\infty)$.} \label{fig:y0}
\end{center}
\end{figure}

The distribution function at freeze-out is obtained by taking the
$\tau/4y \rightarrow \infty$ limit in eqn. (\ref{nofytau}),
therefore we introduce the distribution function at decoupling \be
f_0(y) \equiv n(y;\infty) = 2\Lambda \sqrt{\pi}~
\frac{g_{\frac{5}{2}}(y)}{y^\frac{1}{2}} \,.\label{f0}\ee  This is
the \emph{unperturbed} distribution function that enters in the
Boltzmann-Vlasov equation that determines the evolution of density
and gravitational perturbations, and is displayed in fig.
(\ref{fig:y0}). It is remarkable that for $y \ll 1$ \be f_0(y)
\propto \frac{1}{y^\frac{1}{2}}\,, \label{smallylim}\ee in striking
contrast to the thermal Fermi-Dirac distribution function and to the
one obtained from the (DW) mechanism proposed in
refs.\cite{dw,colombi},
 and closer to that of a (non-condensed)
bosonic massless particle for which the distribution function at
small momentum is $\sim 1/y$.

The divergence of the distribution function $f_0(y)$ for $y  \rightarrow 0$ must be interpreted with
care because we have neglected the build-up of the neutrino population in the quantum kinetic equation,
furthermore we have also neglected terms $\sim m/T(t_f) \sim m/M$ in the derivation. Neglecting the
neutrino population in the kinetic equation requires that $f_0(y) \ll 1$ for all values of $y$, and
neglecting the ratio $m/T(t_f)$ in the frequencies requires that $y \gg m/T(t_f)\sim m/M$. These
constraints imply a range of couplings for which the approximations leading to the final result of the
distribution function are reliable. For example, taking $m \sim ~\mathrm{keV}~;~M \sim 100~\mathrm{GeV}$ neglecting the term $m/T(t_f)\sim m/M$ requires that $y \gg 10^{-8}$. To obtain an estimate of the range
in which the population build-up can be neglected,
it is convenient to write \be \Lambda \simeq 3\,
(Y \times 10^7)^2\,\Bigg( \frac{100}{\overline{g}}\Bigg)^\frac{1}{2}\,\Bigg( \frac{100\,\mathrm{GeV}}{M}\Bigg)
 \,,\label{Yexp}\ee  As discussed above (see also ref.\cite{kusepetra,kuse2})
  taking   the expectation value of the scalar $\langle\chi \rangle \sim M \sim 100\,\mathrm{GeV}$ and
  assuming that the neutrinos obtain their mass via the Yukawa coupling with $m \sim \,\mathrm{keV}$,
   this implies   $Y \sim 10^{-8}$ leading to the illustrative estimate $\Lambda \sim 0.03$.
    The condition for negligible neutrino population becomes \be \frac{\Lambda}{y^\frac{1}{2} } \ll 1 \label{smalpop}\ee
    leading to $y \gg 10^{-4}$ for which we can safely ignore $m/T(t_f)$ \emph{and} the
build-up of the neutrino population. This constraint may be
implemented by introducing an infrared cutoff $y_c \simeq 10^{-4}$
in the $y-$ integrals of the distribution function, however, our
analysis below shows that this is a mild constraint  because of the phase space suppression in these integrals, hence the lower limit
can be safely taken to $y=0$. Values of $\overline{g} > 100$ and $M>
100 \,\mathrm{GeV}$ yield smaller values of $\Lambda$ and larger
region of reliability for $y$.

\section{Constraints from (DM) abundance and (dSphs)  phase space density. }\label{sec:abundance}

The number density \emph{today} of this DM candidate is \be n_0 =
\frac{T^3_0}{2\pi^2} \int^\infty_0 y^2 f_0(y)dy
\label{numberdens}\ee where \be \int_0^\infty y^2 \,f_0(y) dy =
\frac{3\pi}{2}\Lambda \zeta(5)~~;~~\zeta(5) = 1.037\cdots
\label{intfo}\ee

 Since this DM particle is non-relativistic today, its  energy   density       is given by
 \be \rho_{0M}
 = m\,n_0 \,.\label{rho0}\ee
  Entropy conservation\cite{kt} entails that
  \be T_0 = \Bigg(\frac{2}{g_d}\Bigg)^\frac{1}{3}~T_{cmb} \label{Trel}\ee
   where $T_{cmb}=2.348 \times 10^{-4}~\mathrm{eV}$ is the temperature of
   the cosmic microwave background today, and $g_d = \overline{g}$ is the effective
   number of relativistic degrees of freedom at decoupling.

The condition that this DM candidate contributes a fraction $0\leq
\nu_{DM} \leq 1$ to the (DM) density yields the following
\emph{upper} bound on the mass\cite{coldmatter} \be m \leq 2.695
\,\mathrm{eV} \frac{2 \overline{g} \zeta(3)}{g \int^\infty_0
y^2\,f_0(y) dy} \lesssim 1.33
\Bigg(\frac{\overline{g}}{g\,\Lambda}\Bigg)
\,\mathrm{eV}\,,\label{mass}\ee where $g$ is the number of internal
  degrees of freedom of the particle. For a decoupling temperature $\sim 100\, \mathrm{GeV}$ for which\cite{kt}
   $\overline{g} \sim
100$ and $\Lambda \sim 0.01-0.1$ this upper bound is in the
$\mathrm{keV}$ range.

For comparison, for a   fermion relic that decoupled with the
general distribution function (\ref{fdw})   the corresponding upper
bound becomes \be m\leq 3.593 \Big(\frac{ g_d}{\beta\,g}\Big)
\,\mathrm{eV} \label{TFm}\ee For sterile neutrinos produced via the
(DW) mechanism, $\beta$ is adjusted to satisfy this bound with a
given mass and $g_d\sim 30$ (although $g_d$ varies rapidly near
$150\,\mathrm{MeV}$ because of the QCD phase transition or
crossover) and for a thermal relic ($\beta=1$) with  $m \sim
\mathrm{keV}$ it follows that $g_d \gtrsim 500$, namely such thermal
fermion candidate must decouple at a temperature much higher than
the electroweak scale.

As discussed in ref.\cite{coldmatter} a generalized
Tremaine-Gunn\cite{TG} bound that yields a \emph{lower} bound on the
mass is obtained from the coarse grained primordial phase space
density \be \mathcal{D} = \frac{ n(t)}{\langle P^2_f \rangle}
\label{D}\ee where $n(t)$ is the number density of non-relativistic
particles and $\langle P^2_f \rangle$ is the average of the squared
physical momentum in the decoupled distribution function. This
quantity is a Liouville invariant in absence of gravitational
perturbations when the particle has become
non-relativistic\cite{coldmatter}. The   observable quantity is
$\rho/\sigma^3$ where $\rho$ is the mass density and $\sigma$ the
one dimensional velocity dispersion, therefore we define the
primordial phase space density\cite{coldmatter} \be
\frac{\rho_{DM}}{\sigma^3_{DM}} = 3^\frac{3}{2} m^4 \mathcal{D} =
6.611\times 10^8 ~ \mathcal{D}~\Bigg[\frac{m}{\mathrm{keV}} \Bigg]^4
~ \frac{M_\odot/\mathrm{kpc}^3}{(\mathrm{km}/\mathrm{s})^3}\,,
\label{DMPS}\ee where\cite{coldmatter} \be \mathcal{D} =
\frac{g}{2\pi^2} \frac{\Bigg[\int_0^\infty y^2
f_0(y)dy\Bigg]^\frac{5}{2}}{\Bigg[\int_0^\infty y^4
f_0(y)dy\Bigg]^\frac{3}{2}} \,.\label{Den}\ee The distribution
function (\ref{f0}) yields \be \mathcal{D} = \frac{g\Lambda}{2\pi^2}
\frac{\Bigg[\frac{3\pi}{2}~\zeta(5)\Bigg]^\frac{5}{2}}{\Bigg[
\frac{105\pi}{8}~\zeta(7)\Bigg]^\frac{3}{2}} \simeq g\Lambda \times
9.98\times 10^{-3} \label{Df0}\ee whereas for fermions that
decoupled    while relativistic  with the generalized distribution
(\ref{fdw})  $\mathcal{D} \sim g\,\beta \times 1.963\times
10^{-3}$\cite{coldmatter}.

Since the phase space density only diminishes during the merger
process (violent relaxation)\cite{TG,theo} a lower bound on the mass
follows\cite{coldmatter}, \be m \geq
\frac{[62.36~\mathrm{eV}]}{\mathcal{D}^\frac{1}{4}}~ \Bigg[10^{-4}
\frac{\rho}{\sigma^3}
\frac{(\mathrm{km}/\mathrm{s})^3}{M_\odot/\mathrm{kpc}^3}
\Bigg]^\frac{1}{4} \,.\label{lowerboundm}\ee The compilation of
photometric data from (dSphs)\cite{gilmore} yields $ [\cdots
]^\frac{1}{4} \sim 1-2$, taking the middle of this range as an
estimate, the mass of the (DM) particle is bound in the region
between the lower bound (\ref{lowerboundm}) and the upper bound
(\ref{mass}), namely \be
\frac{316~\mathrm{eV}}{(g\Lambda)^\frac{1}{4}} \leq m \leq 1.33
\Bigg(\frac{\overline{g}}{g\,\Lambda}\Bigg) \,\mathrm{eV}
\label{massrange}\ee Taking as representative values
$\overline{g}\sim 100;\Lambda \sim 0.05;g= 2$ yields \be
560~\mathrm{eV} \lesssim m \lesssim 1330 ~
\mathrm{eV}\label{kevrange}\ee constraing  the mass in a fairly
narrow window within the  $\mathrm{keV}$ range. A   similar
conclusion (based on a similar analysis) was obtained in
ref.\cite{petra}.

Thus we see the consistency between the model with scales $\langle
\chi \rangle \sim M \sim 100\,\mathrm{GeV};~ Y \sim 10^{-8};~ m\sim
\mathrm{keV}$ and the constraints from abundance and phase space
density of (dSphs).

\section{Non-relativistic Boltzmann equation: transfer function and power spectrum}\label{sec:TF}
The transfer function is obtained from the solution of the
linearized Boltzmann equation for density and gravitational
perturbations. When the particle has become non-relativistic and for
wavelengths that are well inside the Hubble radius, the
non-relativistic Boltzmann-Vlasov equation describes the evolution
of  these  perturbations. This
equation was used in pioneering work on non-relativistic dark
matter\cite{gilbert}, for neutrinos\cite{bond,bondszalay}, dark
matter perturbations accreted by cosmic strings\cite{bran,bertwatts}
and more recently to study clustering of thermal relic
neutrinos\cite{ringwald}. In all of these previous treatments a
numerical analysis was offered but always with a \emph{thermal}
distribution function  that is   truncated to facilitate the
numerical integration.

Instead, here we follow the analysis of ref.\cite{gildan} and
implement very accurate analytic approximations for the transfer
function that  yield a deeper understanding of the connection
between the decoupled distribution function and the transfer
function, and provide a simple framework to obtain a reliable
assessment of the power spectrum for arbitrary range of parameters
(mass, coupling, etc) and distribution functions.

 To linear order in perturbations
the distribution function of the decoupled particle and the
Newtonian gravitational potential are\cite{peebles,dodelson,bert} \bea f(\vec{p};\vec{x};t) & =
& f_0(p)+ F_1(\vec{p};\vec{x};t) \label{pertdist} \\
\varphi(\vec{x},t)  & =
 & \varphi_0(\vec{x},t)+\varphi_1(\vec{x},t)\,, \label{pertgrav}\eea where
$f_0(p)$ is the unperturbed distribution function of the decoupled
particle, given by (\ref{f0}) and for comparison we will also study
the generalized distribution (\ref{fdw}) with $y=p/T_0$,
$\varphi_0(\vec{x},t)$ is the background gravitational potential
that determines the homogeneous and isotropic
Friedmann-Robertson-Walker metric and $\vec{p},\vec{x}$ are comoving
variables. The reader is referred to
refs.\cite{gilbert,bondszalay,bran,bertwatts,bert,dodelson} for
details on the linearization of the collisionless Boltzmann-Vlasov
equation.

   In conformal time $\tau$ and
in terms of comoving variables $\vec{p},\vec{x}$ it is given by\cite{gilbert,bondszalay,bran} \be
\frac{1}{a} \frac{\partial F_1}{\partial \tau} + \frac{\vec{p}}{m
a^2} \cdot \vec{\nabla}_{\vec{x}} F_1 - m
\vec{\nabla}_{\vec{x}}\varphi_1\cdot \vec{\nabla}_{\vec{p}}f_0 =0
\label{boltzeq}\ee along with Poisson's equation \be
\nabla^2_{\vec{x}} \varphi_1 = \frac{4\pi G m}{a} \int
\frac{d^3p}{(2\pi)^3} F_1(\vec{x},\tau) \,.\label{poissoneq}\ee It
is convenient\cite{gilbert,bran} to introduce a new ``time'' variable
$s$ related to conformal time $\tau$ by \be ds = \frac{d\tau}{a}\,,
\label{svar}\ee and to take spatial Fourier transforms of   $\varphi_1(\vec{x},\tau)$ and   $F_1(\vec{x},\tau)$   obtaining \be \frac{\partial
F_1(\vec{k},\vec{p}\,;s)}{\partial s} + \frac{i \vec{k}\cdot
\vec{p}}{m} F_1(\vec{k},\vec{p}\,;s) -i \vec{k}\cdot
\vec{\nabla}_{\vec{p}}f_0(p) a^2(s) \varphi_1(\vec{k},s) =0\,,
\label{boltzeq2}\ee where \be \varphi_1(\vec{k};s) = -\frac{4\pi G
m}{k^2 a(s)}\int
\frac{d^3p}{(2\pi)^3}F_1(\vec{k},\vec{p};s)\,.\label{poisson2}\ee
The solution of   equation (\ref{boltzeq2}) is
  \be F_1(\vec{k},\vec{p};s) = F_1(\vec{k},\vec{p}\,;s_i) \,e^{-i
\frac{\vec{k}\cdot \vec{p}}{m}(s-s_i)} + i m \vec{k}\cdot
\vec{\nabla}_{\vec{p}} f_0(p) \int^s_{s_i}\,ds' \,e^{-i
\frac{\vec{k}\cdot \vec{p}}{m}(s'-s_i)}a^2(s') \varphi_1(\vec{k},s')\,.
\label{solboltzeq2} \ee The first term on the right hand side is the
free-streaming solution in absence of gravitational perturbations.
The analysis in ref.\cite{gildan} shows that $(p/m)(s-s_i)$ is the
free streaming distance that the particle travels with comoving
velocity $p/m$ from $s_i$ until $s$. Multiplying both sides of eqn. (\ref{solboltzeq2}) by $-4\pi G m/[k^2 a(s)]$,
integrating in $\vec{p}$, and using the relation (\ref{poisson2}),
we obtain \be \varphi_1(\vec{k};s) + i\frac{4\pi G m^2}{k^2 a(s)}
\int \frac{d^3p}{(2\pi)^3} \vec{k}\cdot \vec{\nabla}_{\vec p} f_0(p)
\int^s_{s_i} ds'\,e^{-i \frac{\vec{k}\cdot \vec{p}}{m}(s-s_i)} \,
a^2(s') \varphi_1(\vec{k},s') = -\frac{4\pi G m}{k^2 a(s)} \int
\frac{d^3p}{(2\pi)^3}F_1(\vec{k},\vec{p};s_i) \,e^{-i
\frac{\vec{k}\cdot \vec{p}}{m}(s-s_i)}\,. \label{gil1}\ee The inhomogeneity on the right
hand side of this equation  is determined by the
first term in   (\ref{solboltzeq2}) and describes the
\emph{free streaming} solution of the Boltzmann-equation in absence
of gravitational perturbations.

During matter domination and choosing the initial time at matter-radiation equality $s_i=s_{eq}$, it is found
that\cite{gildan} \be s-s_{eq}=
 \frac{2\,u}{H_{0M}a_{eq}^{\frac{1}{2}}}\,,\label{sofu}\ee where    \be H^2_{0M}
=
 \frac{8\pi G}{3} \rho_{0M} \equiv H^2_0 \Omega_M \,,\ee with $H_0 =   100\,h\, \mathrm{Km}\,\mathrm{sec}^{-1}\,
 \mathrm{Mpc}^{-1}$ is
 the Hubble parameter \emph{today} and $\rho_{0M}$ is the matter density \emph{today}.

 Normalizing the scale factor to unity
\emph{today}, namely $a(0)=1$,  the variable $u$ is
given by\cite{gildan}
 \be   u= 1-\left(\frac{a_{eq}}{a}\right)^{\frac{1}{2}}= 1-\Bigg[\frac{1+z}{1+z_{eq}}\Bigg]^\frac{1}{2}~~;~~ 0\leq u
 \leq 1-a_{eq}^{\frac{1}{2}} \,,\label{varu}\ee and the scale factor in terms of $u$ is given by
 \be  a(u) = \frac{a_{eq}}{(1-u)^2} \,.\label{aofu}\ee  The initial value of the scale factor   at matter-radiation equality is $a_{eq} = 1/  (1+z_{eq})$ with $z_{eq} \simeq 3050$.

  It is convenient to
 introduce the normalized unperturbed distribution function  \be
 \tilde{f}_0(y) = \frac{f_0(y)}{\int_0^\infty y^2 f_0(y) dy}\,,
 \label{tildef0def}\ee which for the non-equilibrium distribution (\ref{f0})  is given
 by \be \tilde{f}_0(y) = \frac{4}{3\sqrt{\pi} \zeta(5)} \frac{g_\frac{5}{2}(y)}{y^\frac{1}{2}}\,. \label{f0tilde}\ee

 Following ref.(\cite{gildan}) we introduce the density perturbation normalized at the initial time \be \delta(\vk,u) = \frac{ \int
\frac{d^3p}{(2\pi)^3} F_1(\vk,\vp\,;s(u)) }{ \int
\frac{d^3p}{(2\pi)^3} F_1(\vk,\vp\,;s_{eq}) } \label{delta}\ee and the
gravitational perturbation normalized at the initial time \be
\Phi(\vk,u)= \frac{\varphi_1(\vk,s)}{\varphi_1(\vk,s_{eq})}
\label{PHI}\ee with the relation (see eqn. (\ref{poisson2}) )\be
{\Phi}(\vk;\,u) = \frac{a_i}{a(u)} {\delta}(\vk ;u) =(1-u)^2
   {\delta}(\vk ;u)\,. \label{fidelta2}\ee The gravitational perturbation $\Phi$    obeys Gilbert's equation\cite{gilbert}
    \be
{\Phi}(\vk,u) - \frac{6 }{\alpha}(1-u)^2 \int^u_0  \Pi[\alpha(u-u')]
\,\frac{ {\Phi}(\vk,u')}{[1-u']^4}
  \,du' = (1-u)^2 I[\vk,u] \,.\label{gil2} \ee where we have introduced\cite{gildan} \be \alpha = \frac{2k \Big(\frac{T_{0}}{m}
\Big)}{ \left[H^2_0 \Omega_M a_{eq}\right]^{\frac{1}{2}} }\simeq 1.18 ~ k \times[\mathrm{kpc}] \Bigg(\frac{100}{\overline{g}}\Bigg)^\frac{1}{3}~\Bigg(\frac{\mathrm{keV}}{m}\Bigg)~\sqrt{1+z_{eq}} \,.
\label{alfa}\ee The non-local kernel  is given by \be
\Pi[z] = \int_0^\infty dy\, y\, \tilde{f_0}(y)  {\sin[yz]} \,,
\label{sigma}\ee and
\be I[\vk,u] =  \frac{\int_0^\infty p^2 d p \, F_1(\vk,p\,;s_{eq}) \,
 j_0\left(\frac{\alpha u p }{T_0}\right)} {\int_0^\infty p^2 d p \, F_1(\vk,p\,;s_{eq})}
  \ee where $j_0(x)=\sin(x)/x$ and   we have assumed that $F_1 $ is   a function of
  $|\vec{p}|$. The inhomogeneity $I[k,u]$ is recognized as the free streaming solution normalized at the initial time, it obeys the initial conditions \be I[\vk,u=0] =1
  ~~;~~\frac{d}{du}I[\vk,u]\Big|_{u=0}=0 \,.\label{iniI}\ee

   The density perturbation $\delta$ obeys \be  {\delta}(\vk,u) - \frac{6}{\alpha}   \int^u_0
\Pi[\alpha(u-u')]\, \frac{ {\delta}(\vk,u')}{[1-u']^2}   \,du' =
I[\vk,u] \,.\label{gildel2} \ee In ref.\cite{gildan} it is proven that $\delta \propto a(t)$ as $u\rightarrow 1$, just like fluid density
perturbations in a matter dominated cosmology.

As analyzed in ref.\cite{gildan}, the transfer function is obtained
from the solution of Gilbert's equation (\ref{gil2})\cite{gildan}.
Normalizing it so that $T(k=0)=1$, it  is given by\cite{gildan} \be
T(k) = {\lim}_{u\rightarrow 1}\frac{\Phi(\vk,u)}{\Phi(\vec{0},u)} =
\frac{5}{3}\, {\Phi(\vk,u=1)} \,. \label{Tofk2}\ee    The final
power spectrum $P_f(k)$ is related to the initial one $P_i(k)$
as\cite{dodelson} \be P_f(k) = T^2(k) P_i(k) \,.
\label{powerspec}\ee If perturbations do not grow or decay
substantially during the prior, radiation dominated phase, $P_i(k)$
is nearly the inflationary primordial power spectrum\cite{dodelson},
which is given  \be P_i(k) = A \Bigg( \frac{k}{k_0} \Bigg)^{n_s}
\label{primopow}\ee where the amplitude $A$ and index $n_s$ are
determined during inflation and $k_0$ is a pivot scale.   The five
years data release from WMAP\cite{WMAP5} yields $n_s \approx 0.96$.


  {\bf Initial conditions:}  The inhomogeneity $I[\vk,u]$ depends on the initial condition $F_1(\vk,p\,;s_{eq})$  which must be specified from the evolution of
    perturbations during the radiation dominated (RD) era up to $t_{eq}$.
     During (RD) the   Boltzmann  equation (\ref{boltzeq}) describes the
    evolution of DM perturbations when the particles are non-relativistic, for the cases under consideration   $T \leq m \sim 1\,\mathrm{keV}$, corresponding to $a  \gtrsim 10^{-4}\,a_{eq}$, and for gravitational perturbations with modes well inside the horizon. During this stage the evolution of the gravitational potential is determined by its coupling to the radiation fluid,  it is given by
     $\varphi_1(k;t)  =3\Phi_p \, j_1(x)/x$ where $\Phi_p$ is the amplitude of primordial perturbations, $j_1$ is the spherical Bessel function, $x=k\eta/\sqrt{3}$ and $\eta$ conformal time\cite{dodelson}.
   $\varphi_1(k;t) $ is  strongly suppressed for wavelengths well inside the horizon and can be considered a small perturbation to the evolution of (DM) perturbations\cite{dodelson}. Therefore the  evolution of non-relativistic perturbations during (RD) is obtained from the solution (\ref{solboltzeq2}) by neglecting the second term that includes the gravitational potential, namely it is given solely by the free-streaming contribution. Consider an initial condition determined early in
   the (RD) era at $s^*$. The free streaming solution of (\ref{solboltzeq2}) during (RD) (neglecting the gravitational potential) is \be \label{FSRD} F_1(\vec{k},\vec{p};s) = F_1(\vec{k},\vec{p}\,;s^*) \,e^{-i
\frac{\vec{k}\cdot \vec{p}}{m}(s-s^*)}~~;~~ s-s^* =  \frac{\ln\left[\frac{a(s)}{a(s^*)}\right]}{\left[H^2_0\,\Omega_M\,a_{eq} \right]^{1/2}}\,  .\ee  Therefore extrapolating this
solution to matter-radiation equality, it follows that the initial condition at $t_{eq}$, namely $F_1(\vec{k},\vec{p}\,;s_{eq})$ is given by (\ref{FSRD}) evaluated at $a(s)=a_{eq}$.

This analysis is akin to the evolution of (CDM) perturbations after decoupling from the radiation fluid studied in ref.\cite{loeb}. An important difference however, is that sterile neutrinos cannot be described as part of a radiation fluid because they do not interact with radiation, leptons or baryons.

We can now follow all the steps described above leading to eqns. (\ref{gil2},\ref{gildel2}) and obtain the same equations but the inhomogeneity $I[k;u]$ replaced by \be I[k;u] \rightarrow  \frac{I[k;u+u_0]}{I[k;u_0]}~~;~~  u_0 = \frac{1}{2}\ln\left[\frac{a_{eq} }{a(s^*)}\right] > 0 \,, \ee  where now \be I[\vk,u] \equiv  \frac{\int_0^\infty p^2 d p \, F_1(\vk,p\,;s^*) \,
j_0\left(\frac{\alpha u p }{T_0}\right)  }{\int_0^\infty p^2 d p \, F_1(\vk,p\,;s^*)}\,,\label{niuI}
  \ee taking $F_1$ to be a function of $|\vec{p}|$. It is clear that  defining the rescaled perturbations \bea \overline{\Phi}(\vk,u) & =  & \Phi(\vk,u) I[k;u_0]  \\ \overline{\delta}(\vk,u) & = &    \delta(\vk,u) I[k;u_0]  \,, \label{redefs}
   \eea these obey  equations (\ref{gil2},\ref{gildel2}) with the inhomogeneity $I[k;u+u_0]$ where $I[k;u]$ is given by (\ref{niuI}). As it will be shown below $I[k;u_0] < 1$ for $u_0 > 0$.  The rescaling of the gravitational and density perturbations by the factor $I[k;u_0]$ reflects the suppression of perturbations by free streaming during the prior (RD) era.

    It now remains to determine $F_1(\vec{k},\vec{p}\,;s^*)$ to finally obtain $I[k;u]$, here we \emph{assume} adiabatic initial conditions corresponding to a temperature
    perturbation, namely \be F_1(\vk,p;s^*) =  \Bigg(T \frac{df_0(p,T)}{dT}\Bigg)\,
\Big(\frac{\Delta T(\vk)}{T}\Big)\,. \label{adiapert} \ee These are
the initial conditions proposed and studied in
refs.\cite{bondszalay,turokWDM}.

   Although a more detailed analysis of the evolution during the radiation era and its impact on the small scale structure will be presented elsewhere,  we can obtain an \emph{upper bound} on the small scale properties of the transfer function by setting $u_0 =0$ ($I[k;0]=1$). Furthermore, since our objective
   is to compare the small scale properties of the transfer function for sterile neutrinos decoupled with the distributions (\ref{f0}) and (\ref{fdw}), it is clear that the distribution (\ref{f0}) leads to a
   \emph{smaller} suppression during (RD) because it favors the small momentum region, therefore yields
   a smaller free streaming velocity and a smaller free streaming length as compared to the distribution (\ref{fdw}).

   \vspace{1mm}

   Hence from the above discussion we conclude the following: i) setting $u_0=0$ leads to an \emph{upper} bound on $T(k)$, ii) the distribution function  (\ref{fdw}) leads to a \emph{larger} free streaming suppression during (RD) as compared to
    the case of the distribution function (\ref{f0}).

\begin{figure}[ht!]
\begin{center}
\includegraphics[height=2in,width=3in,keepaspectratio=true]{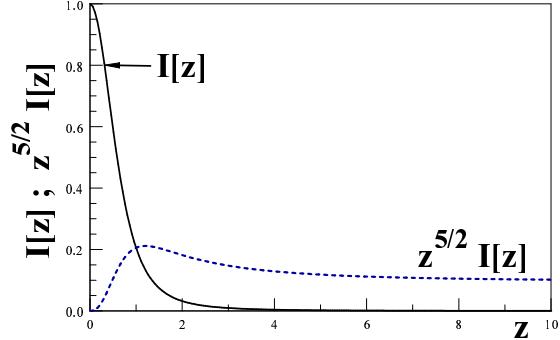}
\caption{ The free streaming solution $I[z]$ eqn. (\ref{Inho}) and $z^\frac{5}{2}I[z]$ vs. $z=\alpha u$.} \label{fig:iofz}
\end{center}
\end{figure}

 With the distribution function
(\ref{f0}) and initial condition (\ref{adiapert}) we find that the normalized free streaming solution is
given by \be I[k,u] = \frac{1}{9\sqrt{2}\zeta(5)}\sum_{n=1}^\infty
\frac{1}{\Big(\rho\,n \Big)^\frac{5}{2}} \Bigg[1+\frac{n}{\rho}
\Bigg]^\frac{1}{2}
\Bigg\{\frac{n}{n+\rho}\Big[1+\frac{3}{\rho^2}\big(n^2-(\alpha
\,u)^2\big) \Big]+ \Big[1+6\frac{n^2}{\rho^2} \Big]\Bigg\}
~~;~~\rho^2 = n^2+(\alpha \,u)^2 \label{Inho}\ee Its asymptotic
behavior   for $\alpha\,u \gg 1$ is \be I[z] \propto  {z
^{-\frac{5}{2}}}~~;~~z=\alpha\,u \,.\label{asyI}\ee  This power law
fall off is strikingly different from that of weakly interacting
massive particles (WIMPs) for which the fall-off by free streaming
is exponential\cite{gildan}, and that for   fermions that decoupled
with generalized distributions (\ref{fdw})  for which the normalized
free streaming solution is the same as for a thermal relativistic
relic (since the suppression factor $\beta$ cancels out) and is also
a power law but with a \emph{faster} fall off $\propto z^{-4}$,
whereas for thermal bosons that decoupled in equilibrium the
fall-off is $\propto z^{-2}$\cite{gildan}. Thus the non-equilibrium distribution
function $f_0(y)$(\ref{f0}) leads to a suppression of the free
streaming solution with a power law intermediate between that of
thermal ultrarelativistic fermions and bosons. Fig. (\ref{fig:iofz})
displays $I[z];z^{5/2}~I[z]$ vs. $z=\alpha\, u$.

It is illustrative to compare the free-streaming solution (\ref{Inho}) to that of a
  fermionic species of the same mass that decoupled at the
same temperature, therefore has the same $\alpha$, and initial condition (\ref{adiapert}) but with the
generalized distribution (\ref{fdw}). The free-streaming solution is
independent of $\beta$ and  is the same as for a thermal
relativistic relic\cite{gildan} \be I_{LTE}[k,u] =
\frac{2}{9\zeta(3)}\,\int^\infty_0 \frac{
y^2\,e^y}{(e^y+1)^2}\,\frac{\sin[y \alpha u]}{\alpha u}\, dy =
\frac{4}{3\zeta(3)} \sum_{n=1}^\infty
\frac{(-1)^{(n+1)}~n}{[n^2+z^2]^2}\Big[1-\frac{4}{3}\frac{z^2}{[n^2+z^2]}
\Big] ~~;~~z=\alpha u \,. \label{ITFD}\ee

 \begin{figure}[ht!]
\begin{center}
\includegraphics[height=2in,width=3in,keepaspectratio=true]{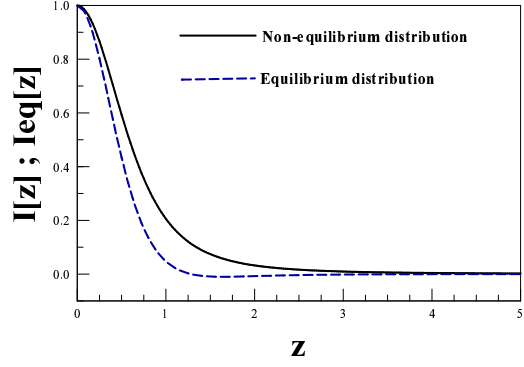}
\caption{ The free streaming solution $I[z]$   for the non-equilibrium distribution (\ref{f0}) eqn. (\ref{Inho})
  (solid line) and for the generalized distribution function   eqn. (\ref{fdw})
(dashed line) vs. $z=\alpha u$. } \label{fig:icompara}
\end{center}
\end{figure}

It is clear that, at least for the initial conditions corresponding
to temperature fluctuations (adiabatic) (\ref{adiapert}), the free-streaming solution
in absence of gravitational perturbations has a slower fall-off in
the case of the non-equilibrium distribution function when compared
to the case of the generalized distribution (\ref{fdw}).

\begin{figure}[ht!]
\begin{center}
\includegraphics[height=2in,width=3in,keepaspectratio=true]{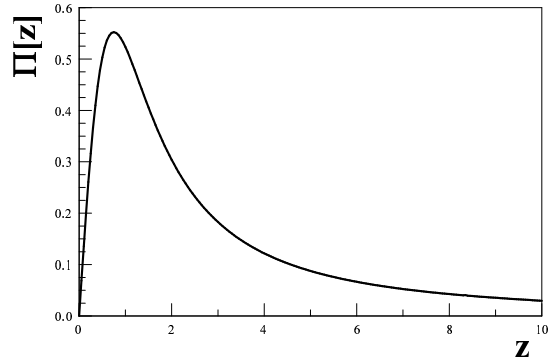}
\caption{ The kernel $\Pi[z]$ vs. $z=\alpha\,(u-u')$  .} \label{fig:piofz}
\end{center}
\end{figure}

  This remarkable difference also emerges in the non-local kernel $\Pi[z]$ in Gilbert's equations for gravitational perturbations(\ref{gil2}) or for density perturbations (\ref{gildel2}). We
find \be \Pi[z]= \frac{\sqrt{2}~z}{\sqrt{3}\,\zeta(5)} \sum_{n=1}^\infty \frac{1}{\Big(\rho\,n\Big)^\frac{5}{2}}\,\Bigg[1+\frac{n}{\rho} \Bigg]^\frac{1}{2}\,\Bigg[\frac{2n+\rho}{n+\rho} \Bigg] ~~;~~\rho=\sqrt{n^2+z^2}\label{Piofz}\ee For $z\sim 0$ $\Pi[z] \propto z$
and asymptotically for $z=\alpha(u-u') \gg 1$ we find $\Pi[z] \propto z^{-3/2}$, in   contrast with the case of thermal fermions for which
the asymptotic behavior is $\Pi[z] \propto z^{-2}$, whereas the asymptotic behavior for thermal bosons
  is found to be $\Pi[z]\propto z^{-1}$\cite{gildan}. Fig.(\ref{fig:piofz}) displays $\Pi[z]$ vs. $z$.

As discussed in detail in ref.\cite{gildan} the longer range of a kernel leads to an enhancement of the transfer function, and to more power at small scales.

\vspace{2mm}

{\textbf{The free streaming wave vector: }}

 The \emph{comoving} free-streaming wave vector is akin to the comoving Jeans wavevector in a fluid but with the
 speed of sound replaced by the velocity dispersion of the decoupled particle, it is  given by \be
  k_{fs}(t)=  k_{fs}(0) \sqrt{a(t)} \label{kfs}\ee and its value \emph{today} is\cite{gildan}
  \be k_{fs}(0)= \Bigg[\frac{3 H^2_{0}\Omega_{M}  }{2 \langle \vec{V}^2 \rangle}\Bigg]^\frac{1}{2}\,,\label{kfs01} \ee
  where  \be \langle \vec{V}^2 \rangle = \frac{\int
d^3 p \Big(\frac{\vec{p}^{~2}}{m^2}\Big) f_0(p)}{\int d^3 p
  f_0(p)} = \Big(\frac{T_{0}}{m}
\Big)^2 \overline{y^2}   \label{V2} \ee is the three dimensional
velocity dispersion  of the
  non-relativistic particles \emph{today} and we introduced \be \overline{y^2} =
\int_0^\infty  dy y^4 \tilde{f}_0(y) \,.\label{overy} \ee Using the
relation (\ref{Trel})  and  $\Omega_M h^2 = 0.105$ for non-baryonic
DM\cite{WMAP5}, leads to\cite{gildan} \be k_{fs}(0)   =
\frac{0.563}{\sqrt{\overline{y^2}}}\,
   \Big( \frac{g_d}{2}\Big)^\frac{1}{3}\Big(\frac{m}{\mathrm{keV}} \Big)\, [\mathrm{kpc}]^{-1}  \,. \label{kfs02}\ee  The variable $\alpha$ defined in eqn. (\ref{alfa}) is related to $k_{fs}(t_i)$ as \be \alpha = \Bigg(\frac{6}{\overline{y}^2}\Bigg)^\frac{1}{2}~\frac{k}{k_{fs}(t_{eq})}\,. \label{alfakfs}\ee
    For the non-equilibrium normalized distribution function (\ref{f0tilde})
    we find \be  \overline{y^2} = \frac{105}{12}~\frac{\zeta(7)}{\zeta(5)} \simeq 8.505 \label{overy2}\ee
     whereas for the generalized distribution (\ref{fdw}) is the same as for thermal relativistic fermions,   \be  \overline{y^2}\big|_{LTE} = 15~\frac{\zeta(5)}{\zeta(3)}
     \simeq 12.939 \,.\label{overyLTE}\ee Therefore the \emph{enhancement} of the non-equilibrium
     distribution function at \emph{small} momenta yields a $\sim 30\%$ reduction in the squared velocity dispersion. Taking the number of relativistic degrees
   of freedom at decoupling $g_d = \overline{g}$ (see eqn. (\ref{Lambdave}) and preceding discussion),
    we find \be k_{fs}(0)=  0.193\, \Big(
\frac{\overline{g}}{2}\Big)^\frac{1}{3}\Big(\frac{m}{\mathrm{keV}} \Big)\,
[\mathrm{kpc}]^{-1} \simeq 0.711 \Big(\frac{\overline{g}}{100}\Big)^\frac{1}{3}~
\Big(\frac{m}{\mathrm{keV}} \Big)\, [\mathrm{kpc}]^{-1} \,   \label{kofsFD}\ee leading to
free streaming wavevector and wavelength at matter-radiation equality\bea   k_{fs}(t_{eq}) &      \simeq  & 0.013 \Big(\frac{\overline{g}}{100}\Big)^\frac{1}{3}~
\Big(\frac{m}{\mathrm{keV}} \Big)\, [\mathrm{kpc}]^{-1} \label{kfseq}\\ {\lambda_{fs}(t_{eq})} &  = & \frac{2\pi}{k_{fs}(t_{eq})} \simeq 488\,
 \Big(\frac{100}{\overline{g}}\Big)^\frac{1}{3}~
\Big(\frac{\mathrm{keV}}{m} \Big)\,{(\mathrm{kpc}) }\,.\label{lambdafseq}\eea

For comparison, the value of $k_{fs}(0)$
 for a fermion that decoupled with the distribution function (\ref{fdw}) is the same as for a relativistic thermal  fermion, given by \cite{gildan} \be  k^{(dw)}_{fs}(0) =  0.157\, \Big(
\frac{g_d}{2}\Big)^\frac{1}{3}\Big(\frac{m}{\mathrm{keV}} \Big)\,
[\mathrm{kpc}]^{-1} \,. \label{kofsLTE}\ee The smaller velocity
dispersion in the case of the non-equilibrium distribution function
yields a $\sim 30\%$ increase in the free streaming wave vector and
consequently a decrease in the free streaming length. But just as
importantly in enhancing $k_{fs}(0)$ is decoupling at high
temperature for a larger value of $g_d$.

For comparison, a sterile neutrino produced via the (DW) mechanism   $T_d \sim 150\,\mathrm{MeV};g_d \sim 30$\cite{dw} yields a free streaming wavevector at matter-radiation equality $\lambda^{(dw)}_{fs}(t_{eq}) \sim 900\,\mathrm{kpc}$.

As discussed in ref.\cite{gildan} an analytic understanding of the
transfer function is obtained by rewriting eqn. (\ref{gildel2}) as a
differential-integral equation. Taking two derivatives with respect
to $u$ and using the original Volterra eqn. (\ref{gildel2}), we
obtain \be \ddot{\td}(\vk,u) - \frac{6\,\td(\vk,u)}{(1-u)^2} +
3\gamma^2 \td(\vk,u) - \int^u_0 du' K[u-u']
\frac{\td(\vk,u')}{(1-u')^2} = \ddot{I}[\vk,u] + 3\gamma^2
I[\vk,u]\,, \label{diffeqn}\ee where dots refer to derivatives with
respect to the variable $u$ and the non-local kernel $K[u-u']$ is
given by \be K[u-u'] = 6 \, \alpha \int_0^\infty y
(\overline{y^2}-y^2 ) \tilde{f}_0(y) \sin[\alpha \,y(u-u')] \,dy \,.
\label{kernel}\ee and $\gamma$ is defined as  \be 3\gamma^2 =
\alpha^2 \overline{y^2} \,. \label{gamma}\ee  From the definitions
(\ref{alfa}) and (\ref{kfs01}) we recognize that the dimensionless
ratio \be \gamma = \frac{
\sqrt{2}\,k}{k_{fs}(t_{eq})}\,.\label{gamadef}\ee It is convenient to
write eqn. (\ref{diffeqn})  as
 \be \ddot{\delta}(\vk,u) -
\frac{6\,\delta(\vk,u)}{(1-u)^2} + 3\gamma^2\,\delta(\vk,u) =
S[\delta;u]\,.\label{differ} \ee The source term is \be S[\delta;u]
= S_0[u]+S_1[\delta;u] \,,\label{source} \ee
where \bea S_0[u] & = & \ddot{I}+3\gamma^2 I \,, \label{S0}\\
S_1[\delta;u] & =&   \int_0^u
du'K[u-u']\,\frac{\delta(\vk,u')}{[1-u']^2} \,.\label{S1} \eea
Passing to cosmic time it is straightforward\cite{gildan} to find
that the left hand side of eqn. (\ref{differ}) is precisely of the
form of the Jeans equation for fluids but with the adiabatic speed of sound
replaced by $\sqrt{\langle \vec{V}^2\rangle}$. However, as discussed in ref.\cite{gildan} the non-local
kernel $K[u-u']$ includes higher order moments of $p^2/m^2$ than those typically kept in the hierarchy
of moment equations\cite{dodelson}.

The two terms in the source $S[\delta;u]$ have very different physical interpretations. The first
term, $S_0$ describes  a ``driving force'' resulting from the  free streaming of the initial perturbation,
the second term $S_1$ is a \emph{correction} to the fluid description and can be interpreted as a
non-local ``pressure'' term. As discussed in ref.\cite{gildan} the second term is negligible in the
long wavelength limit since $K[u-u'] \propto \alpha^4$ for $\alpha \rightarrow 0$,  but becomes important at small scales.
 Furthermore the \emph{memory} of this kernel is determined by the small-$y$ behavior
 of the  distribution function $\tilde{f}_0(y)$\cite{gildan}: larger support at small values of $y$ yields longer range
memory kernels, which enhance the transfer function at small scales. This is a consequence of
the fact that for kernels with longer range, memory of the initial stages of gravitational clustering
persists throughout the evolution leading to a larger contribution to $S_1$\cite{gildan}.

The solution of
(\ref{differ}) can be written exactly in a formal iterative
Fredholm series in which the main ingredients are the mode functions
corresponding to the fundamental regular and irregular solutions of
the homogeneous equation. Defining   \be z= \sqrt{3}\gamma (1-u)
\equiv z_0 (1-u)~~ \,;~~ z_0 = \sqrt{3}{\gamma} \,.\label{zetavar}
\ee these are \bea h_1(z) & = & \left(\frac{3}{z^2}-1\right)\cos(z)+
\frac{3}{z} \sin(z) \,,\label{h1}\\ h_2(z) & = &
\left(\frac{3}{z^2}-1\right)\sin(z)- \frac{3}{z} \cos(z)\,,
\label{h2}\eea with the following asymptotic behavior as
$z\rightarrow 0$ ($u \rightarrow 1$) \bea h_1(z)\rightarrow
\frac{3}{z^2} ~~;~~h_2(z) \rightarrow \frac{z^3}{15}
\,.\label{hsasy} \eea

We emphasize  that the kernels $\Pi[z];K[z]$ \emph{do not depend on
the overall normalization of the distribution function}. This is
important because Dodelson-Widrow-type distribution functions
(\ref{fdw}) (see also ref.\cite{colombi}) are of the form \be
f_{dw}(y) = \frac{\beta}{e^y +1}\,, \label{fdw2}\ee where $0 \leq
\beta \leq 1$ is a \emph{suppression} factor. For these distribution
functions the normalized counterpart \be \tilde{f}_{dw}(y) =
\frac{f_{dw}(y)}{\int_0^\infty y^2\,f_{dw}(y)dy} =
\frac{2}{3\zeta(3)} \frac{1}{e^y+1}\, \label{FDnor}\ee is the same
as that for a relativistic fermionic thermal relic. The suppression
factor $\beta$ only enters in the abundance and the primordial phase
space density. Therefore for sterile neutrinos produced by the
(DW) mechanism\cite{dw,colombi} or for general
distribution functions of the form (\ref{fdw}),  the kernels in
 Gilbert's equation (\ref{gil2}) or in the equivalent equation (\ref{diffeqn}) and the
 free streaming solution in absence of gravitational perturbations $I[k,u]$ are the
\emph{same as for the case of a fermionic relativistic thermal
relic}. This results in that the transfer function and power
spectrum for sterile neutrinos produced by non-resonant mixing \emph{are
the same as that for relativistic thermal fermions } (see below).
This observation is relevant in view of the stringent constraints from the analysis in
ref.\cite{lyman2}. The results in this reference are based on
hydrodynamical simulations that assume a sterile neutrino
distribution function of the form (\ref{fdw2}). The overall
multiplicative   normalization \emph{does} not affect the non-local
kernel $K[z]$ and as a result, nor does it affect the transfer
function and the power spectrum (see below). This kernel features
the asymptotic behavior $\propto 1/z^2$ for any distribution of the
form (\ref{fdw}) as  used in ref.\cite{lyman2} for the simulations,
whereas for the distribution function (\ref{f0}) $K[z] \propto
1/z^{\frac{3}{2}}$. Namely, the kernel falls-off \emph{slower} than in the
case of the Dodelson-Widrow production mechanism leading to a longer range of memory
of gravitational clustering. The analysis in
ref.\cite{gildan} indicates that this feature, slower fall-off and longer memory range translates in an
enhanced transfer function at small scales. This behavior will be
confirmed below.

\subsection{ The transfer function and power spectrum}

 From the Fredholm series solution for $\delta$ the
\emph{exact} transfer function is given by\cite{gildan} \be T(k)=
\frac{10}{  \sqrt{3}\,\gamma^3 } \int^1_0 h_2(u')
\Bigg[\frac{I[\vk,u']}{(1-u')^2} +\frac{ S_1[\td;u']}{6} \Bigg]\,du'
\,,\label{Tint}\ee where $\td$ in $S_1$ is the Fredholm solution of
the integral equation (\ref{differ}). As shown in detail in
ref.\cite{gildan} a remarkably accurate approximation to the full
transfer function is obtained from the first two terms in the
Fredholm series  \be T(k)\simeq T_{B}(k) + T_{(2)}(k) = \frac{10}{
\sqrt{3}\,\gamma^3 } \int^1_0 h_2(u') \Bigg[
\frac{I[\vk,u']}{(1-u')^2} + \frac{ S_1[\td^{(1)};u']}{6} \Bigg] du'
\, \label{Tint2ndord}\ee where the first, Born term is  given by \be
T_B(k) = \frac{10}{  \sqrt{3}\gamma^3 } \int^1_0 h_2(u')
\frac{I[\vk,u']}{(1-u')^2} du'\label{Tborn} \ee and the second order
correction is given by  \be T_{(2)}(k)=
  \frac{5}{3\sqrt{3}\,\gamma^3 } \int^1_0 h_2(u')  S_1[\td^{(1)};u']
 \,du' \,, \label{T2ndord}\ee   where
 $\delta^{(1)}(k;u) $ is given by  \be \td^{(1)}(\vk,u) = I[\vk,u] +
\frac{6}{\sqrt{3}\,\gamma} \int_0^u
\left[h_1(u)h_2(u')-h_2(u)h_1(u')\right]\frac{I[\vk,u']}{(1-u')^2}
  du' \,. \label{td1}\ee

Since the free streaming solution $I[k;u]$, the kernels and the mode functions are all
functions of $\alpha$ (see eqn. (\ref{alfa})), it follows that $T(k)$ is   a function
of $\alpha$. Therefore it proves convenient to relate the comoving wavelength $\lambda = 2\pi/k$ to
$\alpha$ in order to establish the scales that enter in $T(k)$,  namely \be
\frac{\lambda}{(\mathrm{kpc})} \simeq \frac{409}{\alpha} \Bigg(
\frac{100}{g_d}\Bigg)^\frac{1}{3}\, \Bigg(
\frac{\mathrm{keV}}{m}\Bigg) \label{lamalfa}\ee where $g_d
=\overline{g}$ for sterile neutrinos with the non-equilibrium
distribution function (\ref{f0}).

  The contribution from $T_{(2)}(k)$ is the first correction beyond the fluid
  approximation and  includes the memory of
  the initial conditions and gravitational clustering. This correction is negligible in the long wavelength
  limit $\alpha \rightarrow 0$ but  becomes
  important at short scales\cite{gildan}.

\begin{figure}[ht!]
\begin{center}
\includegraphics[height=2in,width=3in,keepaspectratio=true]{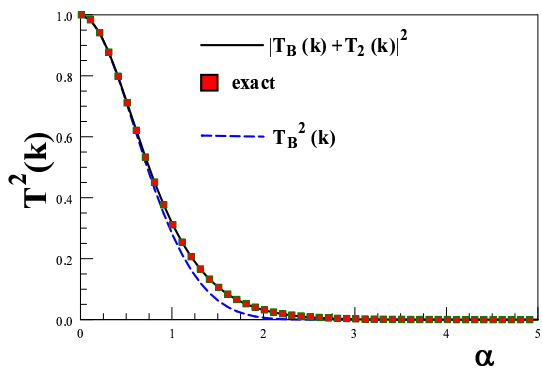}
\caption{ Comparison between the exact (red squares) solution, the Born approximation
and the second order improvement.} \label{fig:exactaprox}
\end{center}
\end{figure}

  Fig. (\ref{fig:exactaprox}) displays the Born term $T^2_B(k)$, the second order
  corrected $(T_B(k)+T_{(2)}(k))^2$ and the exact $T^2(k)$ obtained from the numerical integration of
  Gilbert's equation (\ref{gil2}) with the distribution function (\ref{f0}).
  This figure shows that: (a) the second order correction becomes
  important at small scales  $\alpha >1$ and (b) that the Born plus second order correction approximation
  to the transfer function is a remarkably accurate. The outstanding accuracy of the second order
  approximation was also pointed out in ref.\cite{gildan} for thermal relics.

\begin{figure}[ht!]
\begin{center}
\includegraphics[height=2in,width=3in,keepaspectratio=true]{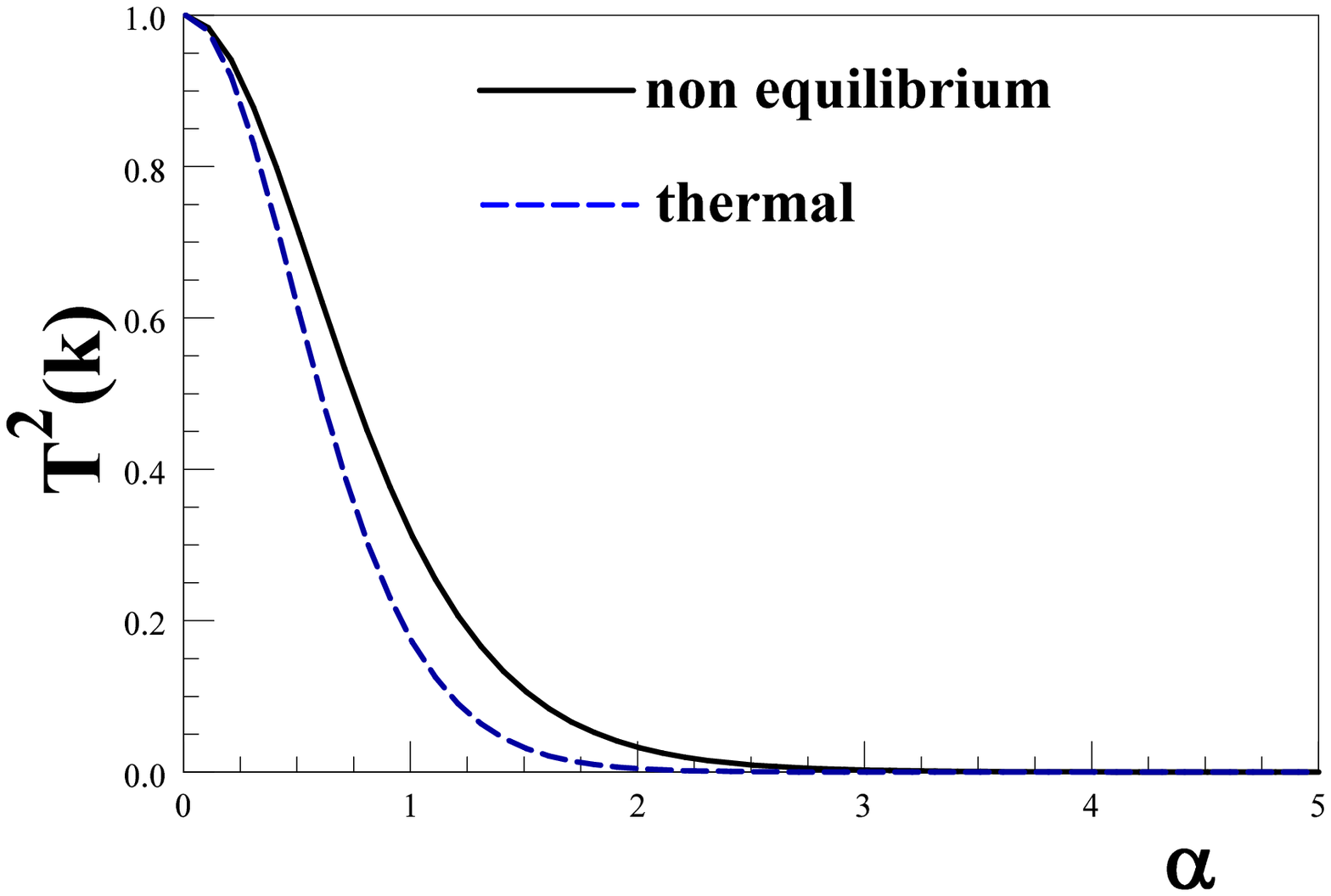}
\includegraphics[height=2in,width=3in,keepaspectratio=true]{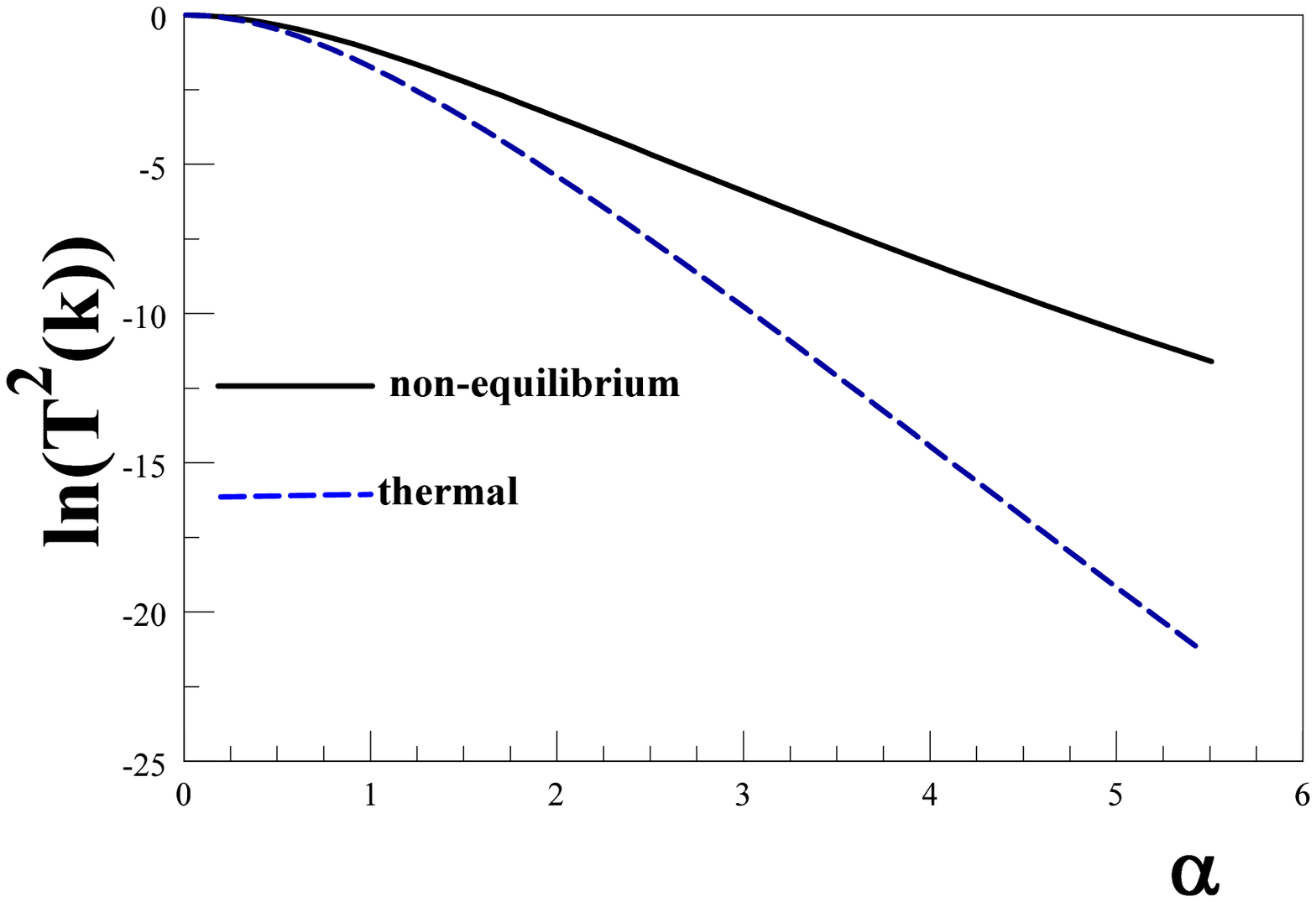}
\caption{ Left panel: comparison between $T^2(k)$ for thermal fermions and sterile neutrinos
decoupled out of equilibrium, right panel: comparison between $\ln(T^2(k))$ for both cases.  } \label{fig:compara}
\end{center}
\end{figure}

Fig. (\ref{fig:compara}) compares $T^2(k)$ for fermions that
decoupled with the generalized distribution function (\ref{fdw}),
because the normalized distribution is the same as the case of
thermal relativistic relics we refer to this case as \emph{thermal},
and with the non-equilibrium distribution function (\ref{f0})
(non-equilibrium). The right panel displays $\ln(T^2(k))$ for both
cases to make explicit the \emph{enhancement} of the transfer
function for the non-equilibrium case at small scales $\alpha > 1$.
There are two different sources of this small scale enhancement: i)
the initial condition has a slower fall-off with $k$ in the
non-equilibrium case and ii) the kernels $\Pi[z]$ and consequently
$K[z]$ also have a slower fall-off with $k$ for large $k$. Both
aspects are a consequence of the enhancement of the distribution
function for small $y$.

   Distribution functions that favor the
  small momentum region yield memory kernels that fall off slower an
  enhance the transfer function and power spectrum at small
  scales\cite{gildan}. This small scale (large $\alpha$) enhancement is clearly
  exhibited in the comparison in fig. (\ref{fig:compara}).

  Although the simple analytic approximation is easily to study numerically, it is
  illuminating to provide fitting formulae for $T(k)$ in different ranges of scales.
  For large scales $k \ll k_{fs}(t_{eq})$ equation (\ref{diffeqn}) can be solved in perturbation
  theory in $\gamma$ (or $\alpha$) and the contribution from the non-local kernel $K$ can be neglected
   to leading order in $\alpha$, since in the long wavelength limit $K \propto \alpha^4$. In the long-wavelength limit $k\ll k_{fs}(t_{eq})$ we find that $T(k)$ is approximated by
  \be T(k) \sim 1-C\bigg(\frac{k}{k_{fs}(t_{eq})}\bigg)^2 + \cdots \,,\label{smalk}\ee with $C \sim \mathrm{O}(1)$ and depends on the  distribution function\cite{gildan}. Of more interest is the small scale behavior for $ k \geq k_{fs}(t_{eq})$, because at small scales the contribution from the non-local kernel which is a correction to the \emph{fluid description} that includes
  memory of gravitational clustering  becomes important.   Figure (\ref{fig:compara})
  shows that $T(k)$ can be approximated by an exponential for $\alpha \gtrsim 0.8$. A numerical
  analysis yields      \be T(k) \simeq 1.902 \, e^{-k/k_{fs}(t_{eq})} ~~;~~k \geq k_{fs}(t_{eq})  \,,\label{Tsmalk}\ee in the range $ 0.8 \lesssim \alpha \leq 6 $, where \be k_{fs}(t_{eq}) \simeq 0.013\, \Big(\frac{\overline{g}}{100}\Big)^\frac{1}{3}~
\Big(\frac{m}{\mathrm{keV}} \Big)\, [\mathrm{kpc}]^{-1} \,.\ee The fit
(\ref{smalk})  is better than $2\%$ in this range. For $\overline{g} \sim 100$ and $m \sim 1 \,\mathrm{keV}$   the fit
(\ref{Tsmalk}) accurately describes the transfer function in the
range \be 65 \,\mathrm{kpc} \lesssim \lambda \lesssim
500\,\mathrm{kpc}\,,\label{lambdarange}\ee approximately from the
scale of clusters of galaxies to that of galaxies. Within this range
$ 10^{-5} \leq T(k) \leq 0.7   $.

The excellent fit  (\ref{Tsmalk}) is   different from the often
quoted numerical fit by Bardeen \emph{et.al.}\cite{BBKS}.

  \vspace{2mm}

  {\bf Power spectrum:}

Since $T(k)$ is a function of the combination $\alpha$ given by eqn.
(\ref{alfa}) it is convenient to write the power spectrum \be P(k) =
A \Bigg( \frac{k}{k_0}\Bigg)^{n_s} T^2(k) \equiv B \, \alpha^{n_s}~
T^2(k)\,. \label{powalfa}\ee

\begin{figure}[ht!]
\begin{center}
\includegraphics[height=2in,width=3in,keepaspectratio=true]{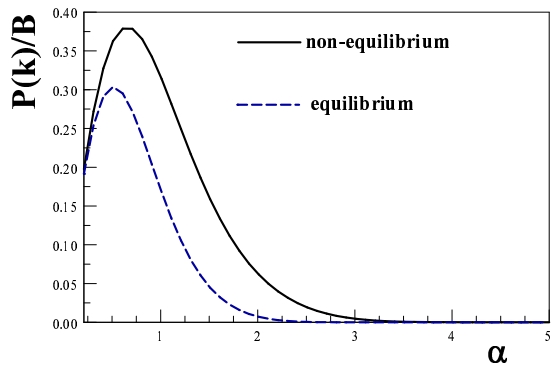}
\caption{ The spectrum $P(k)/B$ vs. $\alpha$ for the non-equilibrium distribution function (solid line)
 compared to a thermal fermion relic (dashed line).} \label{fig:power}
\end{center}
\end{figure}

Fig. (\ref{fig:power}) displays $P(k)/B$ vs. $\alpha$   for sterile
neutrinos decoupled with (\ref{fdw}) (equilibrium) and (\ref{f0})
  (non-equilibrium) as a function of $\alpha$. The
figure reveals that at large scales $\alpha \ll 1$ both feature the
same power spectrum (which is the same as that for cold dark matter)
but a substantial difference emerges at small scales. The
non-equilibrium distribution function (\ref{f0}) yields an enhanced
transfer function, hence more power at small scales. This is a
consequence of the fact that the non-equilibrium distribution
function favors smaller values of the momenta (small values of $y$),
leading to smaller velocity dispersion  hence \emph{effectively
colder } particles, smaller free streaming length, but more
importantly a memory kernel of longer range. This feature results in
that memory of the gravitational clustering ``lingers'' longer and
the initial value of the gravitational potential influences the
process of gravitational clustering during a longer period of
time\cite{gildan} leading to an enhancement of the transfer function
and the power spectrum at small scales.

Fig. (\ref{fig:compara}) shows that the suppression scale of
$T^2(k)$ for   relics that decoupled with   (\ref{fdw}) is at
$k \simeq k_{fs}(t_{eq})$. For sterile neutrinos produced via
the Dodelson-Widrow mechanism at a temperature $T_d \sim
150\,\mathrm{MeV}$, with $g_d \sim 30$ and $m = 1\,\mathrm{keV}$
this scale corresponds to a comoving wavelength $\lambda^{(dw)}_{fs}(t_{eq}) \sim 0.9
\,\mathrm{Mpc}$, whereas for a $\mathrm{keV}$ sterile neutrino
produced in the model under consideration that decoupled with
(\ref{f0}) with $\overline{g} \sim 100$ this scale corresponds to
  a comoving wavelength $\lambda_{fs}(t_{eq}) \sim 0.49
\,\mathrm{kpc}$. At smaller scales, for $\alpha \gg 1$ the
difference in $T^2(k)$ for thermal relics and the non-equilibrium
distribution becomes more dramatic as shown in the right panel of
fig.(\ref{fig:compara}).

The small scale enhancement of $T(k)$ is a consequence of the small $y$ behavior of the distribution
function, which translates into a longer range kernel $K[z]$. Thus it is clear from these figures that
the non-equilibrium distribution function, combined with the higher decoupling temperature, namely the \emph{colder} behavior (smaller velocity dispersion) yield a substantial enhancement of power at
small scales as compared to either thermal relics or to sterile neutrinos produced via non-resonant mixing
with active neutrinos (namely \emph{ a la} Dodelson-Widrow).

For scales $\lambda \gg  1 \mathrm{Mpc}$, namely $\alpha \ll 0.4$ the transfer functions and power
spectra of sterile neutrinos produced by scalar decay, the (DW) mechanism, relativistic thermal
relics or (CDM)  are essentially indistinguishable. The non-equilibrium case begins to feature larger
power than (DW) and relativistic thermal relics for $\alpha > 0.5$, namely for scales $\lambda \lesssim 0.8 \,\mathrm{Mpc}$ and becomes substantially \emph{larger} than either of these cases for small scales $\lambda \lesssim 490\,\mathrm{kpc}$.

\section{Conclusions and further questions}

In this article we have implemented a program that begins with the microphysics of production and
decoupling of a dark matter particle candidate, constrains the mass and couplings from the observed
DM abundance and phase space density of DM dominated satellite galaxies (dSphs) and obtains the
DM transfer function and power spectrum by solving the Boltzmann-Vlasov equation for density and gravitational perturbations during matter domination.

The model studied is a phenomenologically appealing extension of the minimal standard model proposed in
references\cite{shapo,kusepetra,petra,kuse2} in which sterile neutrinos are produced by the decay of
a gauge singlet scalar. With the scale of the expectation value and mass of this scalar $\sim 100\,\mathrm{GeV}$ a consistent description of $\sim \mathrm{keV}$ sterile neutrinos decoupled strongly out of equilibrium at a decoupling temperature $\sim 100\,\mathrm{GeV}$ emerges and satisfies the DM abundance
and phase space constraints from (dSphs).

The distribution function after decoupling for sterile neutrinos produced by gauge singlet decay features
a strong enhancement at small comoving momentum $\propto 1/\sqrt{p}$ in contrast to sterile neutrinos
produced via non-resonant mixing with active neutrinos (\emph{a la} Dodelson-Widrow) for which   the distribution function is
that  of ultrarelativistic thermal fermions multiplied by a suppression factor. Such distribution
function was used in the hydrodynamical simulations performed in ref.\cite{lyman2} to analyze the
Lyman-$\alpha$ forest data.

We have implemented an accurate analytic approximation to the solution of the Boltzmann-Vlasov equation and the
transfer function introduced in ref.\cite{gildan} and  obtained the power spectrum. This
approximation allows to identify which features of the distribution function determine the small scale behavior of the transfer function. Distribution functions that favor small (comoving) momentum lead to
longer range memory of gravitational clustering and slower fall-off of the free streaming solution. Both
features lead to \emph{small scale enhancement} of the transfer function and power spectrum.

 We compare
the transfer function and power spectrum in the case of sterile
neutrinos produced by gauge singlet decay and by non-resonant mixing
with active neutrinos, and find that the former is substantially
enhanced over  latter at small scales. The transfer function and
power spectrum of sterile neutrinos produced via non-resonant mixing
is the \emph{same} as that for fermionic thermal relics. The
suppression factor in the distribution (see eqn. \ref{fdw}) modifies
the abundance an primordial phase space densities but \emph{not} the
transfer function or power spectrum.

While at large scales $\lambda \gg 1\,\mathrm{Mpc}$ the transfer
function in both cases are nearly indistinguishable and the same as
the (CDM) case, the power spectrum for $m\sim \mathrm{keV}$ (DW)
sterile neutrinos produced by non-resonant mixing is suppressed
below a scale $\lambda \lesssim 900 \,\mathrm{kpc}$ whereas the
transfer function for sterile neutrinos produced via scalar decay is
suppressed below a scale $\lambda \lesssim 488 \,\mathrm{kpc}$ and
substantially enhanced at smaller scales when compared to the (DW)
case.

We find the simple fits to $T(k)$ in the limits of large and small scales. For large scales we
find  \be T(k) \sim 1-C\bigg(\frac{k}{k_{fs}(t_{eq})}\bigg)^2 + \cdots ~~;~~ {k} \ll k_{fs}(t_{eq})\,, \ee with $C \sim \mathrm{O}(1)$. In this long wavelength limit the fluid description is valid and the contribution from the memory kernel is subleading.

At small scales the corrections to the fluid description in terms of the non-local kernel that includes memory of gravitational clustering  becomes important, in the small scale regime ${k} \geq k_{fs}(t_{eq})$  a   simple and   accurate numerical fit yields:

\be T(k) \simeq 1.902\,e^{-k/k_{fs}(t_{eq})}\,, \ee where $k_{fs}(t_{eq})$ is
the free streaming wave vector at matter-radiation equality. For  a sterile neutrino with $m\sim 1\,\mathrm{keV}$
decoupling at $T_d\sim 100 \,\mathrm{GeV}$ we find \be k_{fs}(t_{eq}) \simeq 0.013/\mathrm{kpc}\,.\ee  This
  fit is remarkably accurate in the wide range of scales $60\,\mathrm{kpc} \lesssim \lambda \lesssim 500\,\mathrm{kpc} $ and is   different from the often quoted result of ref.\cite{BBKS}.


We  have given arguments that show that the results presented above are an   \emph{upper bound} to the small scale properties of $T(k)$, since the evolution of WDM perturbations during   (RD) leads to
further suppression of $T(k)$ with a \emph{larger} suppression for the case of sterile neutrinos with distribution functions of the form (\ref{fdw}) as compared to those with (\ref{f0}). This is because the
distribution function (\ref{f0}) favors the small momentum region leading to shorter free streaming lengths and larger free streaming wavevectors, allowing more power at small scales.  A  more  detailed analysis of the initial conditions obtained by including the evolution during the (RD) will be reported elsewhere.


The substantial difference between the suppression scales in the
transfer function and power spectrum at small scales between sterile
neutrinos produced by gauge singlet decay and those produced by the
(DW) mechanism suggest that sterile neutrinos produced by scalar
decay \emph{may } relieve the tension between the constraints from X-ray
\cite{xray} and Lyman-$\alpha$ forest data\cite{lyman,lyman2}.

In order to assess whether sterile neutrinos produced by scalar
decay may explain the cored profiles of (dSphs), a full N-body
simulation with the power spectrum obtained above must be carried
out.

Whereas in this article we have focused on the production mechanism from gauge singlet decay, the
model includes sterile-active mixing via a see-saw (Majorana) mass matrix. Therefore there is also
a complementary mechanism of sterile neutrino production via active-sterile mixing akin to the (DW) mechanism,
 which is effective at a much lower temperature $\sim 150\,\mathrm{MeV}$. The wide separation
of decoupling scales, $\sim 100\,\mathrm{GeV}$ for scalar decay, vs.
$\sim 150\,\mathrm{MeV}$ for (DW) suggests that once the
distribution function has been established after decoupling at the
higher temperature, the non-equilibrium effects at the lower scale
may not modify the small momentum region significantly. We
conjecture this to be the case because the wide separation of scales
\emph{suggests} that the distribution function obtained from scalar
decay, may be taken as an \emph{initial condition} for the kinetics
of production via active-sterile mixing, in which case for small
mixing angle and neglecting again the build up of the population,
the final distribution function would be the \emph{sum} of
(\ref{f0}) and (\ref{fdw}).

We have not included  this possibility in this article, postponing
the more detailed kinetic description, a study of the consequences
and different initial conditions to a forthcoming article.

In obtaining the transfer function and power spectrum by solving the
Boltzmann-Vlasov equation for (DM) and gravitational perturbations
during matter domination, we have neglected the contribution from
baryons and photons. Although these will only affect the results for
$T(k)$ and $P(k)$ at the few percent level, as discussed in the
introduction, a precise assessment of $P(k)$ to few percent accuracy
requires solving the full set of Boltzmann-equations by modifying
publicly available codes.  The study presented in this article
provides a reliable preliminary assessment of   (DM) candidates,
allows a systematic comparison  and highlights the important
small-scale aspects, perhaps eventually justifying the more
numerically demanding task of solving the full set of coupled Boltzmann equations for photons,
baryons, gravitational and (DM) perturbations.

\acknowledgements{The author thanks  Alex Kusenko for stimulating
discussions. He acknowledges support
 from the U.S. National Science Foundation through grant No:  PHY-0553418.}

\appendix

\section{The quantum kinetic equation} With the Yukawa interaction
$\mathcal{L}_I = Y\,\chi \,\overline{\nu}\,\nu$ and $\chi$ a scalar
field with mass $M$ and $\nu$ either a Dirac or Majorana fermion
field of mass $m$ the quantum kinetic equation is obtained just as
in Minkowski space time by obtaining the total transition
probabilities per unit time of the decay and inverse decay
processes.

\begin{figure}[ht!]
\begin{center}
\includegraphics[height=2in,width=3in,keepaspectratio=true]{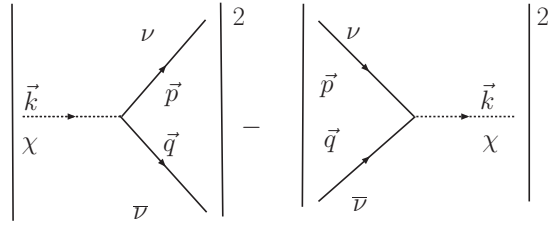}
\caption{The ``gain'' and ``loss'' contributions to the quantum
kinetic equation from the decay $\chi \rightarrow
\overline{\nu}\,\nu $ and inverse decay $\overline{\nu}\,\nu
\rightarrow \chi$, to lowest order in the Yukawa coupling.}
\label{fig:rate}
\end{center}
\end{figure}

The quantum kinetic equation for the neutrino population $n(p;\,t)$
(and similarly for the antineutrino) is  of the usual form \be
\frac{d\,n(p;\,t)}{d t} = \mathrm{Gain}-\mathrm{Loss} \ee where the
gain and loss terms are obtained from the corresponding transition
probabilities $|\mathcal{M}_{fi}|^2$. The processes are depicted in
fig. (\ref{fig:rate}).

The gain term is obtained from the decay reaction $\chi \rightarrow
\overline{\nu} \, \nu$ depicted in the first term in fig.
(\ref{fig:rate}) corresponding to an initial state with $N_k$ quanta
of the scalar field $\chi$ and $n_{p,s},\overline{n}_{q,s'}$ quanta
of  neutrinos and antineutrinos respectively with
$N_k-1,n_{p,s}+1,\overline{n}_{q,s'}+1$ quanta in the final state
respectively. The corresponding Fock states are \be |i\rangle =
|N_k,n_{p,s},\overline{n}_{q,s'}\rangle  ~~;~~|f\rangle =
|N_k-1,n_{p,s}+1,\overline{n}_{q,s'}+1\rangle  \label{inistate} \ee
The loss term is obtained from the inverse decay reaction
$\overline{\nu}\,\nu \rightarrow \chi$  corresponding to an initial
state with $N_k$ quanta of the scalar field $\chi$ and
$n_{p,s},\overline{n}_{q,s'}$ quanta of  neutrinos  and
antineutrinos respectively with
$N_k=1,n_{p,s}-1,\overline{n}_{q,s'}-1$ quanta in the final state
respectively. The corresponding Fock states are \be |i\rangle =
|N_k,n_{p,s},\overline{n}_{q,s'}\rangle  ~~;~~|f\rangle =
|N_k+1,n_{p,s}-1,\overline{n}_{q,s'}-1\rangle  \label{finstate} \ee
The calculation of the matrix elements is standard, the fields are
quantized in a volume $V$ in terms of Fock creation-annihilation
operators, with the corresponding spinor solutions for the neutrino
fields. To lowest order in the Yukawa coupling, \be
\mathcal{M}_{fi}\Big|_{\mathrm{gain}} = -i \frac{Y}{\sqrt{V}}
\frac{\delta_{\vec{k},\vec{p}+\vec{q}}}{\sqrt{2\Omega_k}}\sqrt{N_k}\sqrt{1-n_p}\sqrt{1-\overline{n}_q}~~
\overline{U}_{\alpha}(\vec{p},s)V_\alpha(\vec{q},s')\,
(2\pi)\delta(\Omega_k-\omega_p-\omega_q) \label{Mfigain}\ee
similarly, for the loss term we obtain, \be
\mathcal{M}_{fi}\Big|_{\mathrm{loss}} = -i \frac{Y}{\sqrt{V}}
\frac{\delta_{\vec{k},\vec{p}+\vec{q}}}{\sqrt{2\Omega_k}}\sqrt{1+N_k}\sqrt{n_p}\sqrt{\overline{n}_q}~~
\overline{V}_{\alpha}(\vec{q},s')U_\alpha(\vec{p},s)\,
(2\pi)\delta(\Omega_k-\omega_p-\omega_q) \label{Mfiloss}\ee where
the spinors have been normalized to one and the frequencies \be
\Omega_k = \sqrt{k^2+M^2}~~;~~\omega_p= \sqrt{p^2+m^2}\,.
\label{freqs}\ee  Summing the respective $|\mathcal{M}_{fi}|^2$ over
$\vec{k},\vec{q},s,s'$  and taking the infinite volume limit we
obtain the total transition probability for the gain term \be
\sum_{\vec{k},\vec{q},s,s'}~|\mathcal{M}_{fi}|^2\Big|_{\mathrm{gain}}
= \mathrm{T}~\frac{Y^2}{4\pi}\int d^3q
\frac{\delta(\Omega_k-\omega_p-\omega_q)}{
\Omega_{\vec{p}+\vec{q}}\,\omega_p\,\omega_q}~\big[\omega_p\omega_q-\vec{p}\cdot\vec{q}-m^2
\big] \Big[N_{\vec{p}+\vec{q}}\,(1-n_p)\,(1-\overline{n}_q) \Big]
\label{Gamagain}\ee where $\mathrm{T}$ is the total reaction time.
For the loss term we find the same expression but with the
replacement $N_{\vec{p}+\vec{q}}\rightarrow
1+N_{\vec{p}+\vec{q}}~,~(1-n_p)\rightarrow
n_p~,~(1-\overline{n}_q)\rightarrow  \overline{n}_q $. Carrying out
the angular integral using the energy-conserving delta function we
obtain the final expression for the neutrino production rate \be
\frac{d\,n(p;t)}{d\,t} = \frac{1}{\mathrm{T}} \Big[
\sum_{\vec{k},\vec{q},s,s'}~|\mathcal{M}_{fi}|^2\Big|_{\mathrm{gain}}
-
\sum_{\vec{k},\vec{q},s,s'}~|\mathcal{M}_{fi}|^2\Big|_{\mathrm{loss}}
\Big] = \frac{Y^2}{8\pi}
\frac{\Big(1-\frac{4m^2}{M^2}\Big)}{p\omega_p} \int^{q_+}_{q_-}
\frac{qdq}{\omega_q}
\Big[N_{\vec{p}+\vec{q}}~(1-n_p)~(1-\overline{n}_q)-(1+N_{\vec{p}+\vec{q}})~n_p
~\overline{n}_q \Big]   \label{rateeqn} \ee where $q_{\pm}$ are the
roots of the equations \be \Big[(p\pm
q_{\pm})^2+M^2\Big]^\frac{1}{2} = \omega_{q_{\pm}}+\omega_p
\label{roots}\ee As discussed in section (\ref{sec:model}) the
relevant limit is $M\gg m$. In this limit $M\gg m$ we find these
roots to be \be q_{\pm} = \frac{M^2}{2m^2} \big(\omega_p \pm p\big)
\label{qs}\ee

The extension to the cosmological case replaces the momenta by the
physical momenta \be p \rightarrow P_f(t)= \frac{p}{a(t)}\,,
\label{physmom}\ee and for Majorana neutrinos $n=\overline{n}$.

\end{document}